\newcommand{\hamilt}{{\hat{\mathcal{H}}}}
\newcommand{\spop}{\hat{\mathbf{S}}}
\newcommand{\spin}{\hat{S}}

\newcommand{\ket}[1]{\left| #1 \right\rangle}

\newcommand{\CCoB}{Cs$_2$CoBr$_4$}
\newcommand{\CCoC}{Cs$_2$CoCl$_4$}
\newcommand{\CCuB}{Cs$_2$CuBr$_4$}
\newcommand{\CCuC}{Cs$_2$CuCl$_4$}

\documentclass[aps,pra,reprint,showpacs,longbibliography, citeonline]{revtex4-1}
\usepackage{amsmath,amssymb,bm}
\usepackage{graphicx}
\usepackage{xspace}
\usepackage{latexsym}
\usepackage{color}
\usepackage{hyperref}
\usepackage{placeins}

\begin{document}
\title{Magnetization plateaux cascade in the frustrated quantum antiferromagnet Cs$_2$CoBr$_4$}

\author{K.~Yu.~Povarov}
        \email{povarovk@phys.ethz.ch}
        \affiliation{Laboratory for Solid State Physics, ETH Z\"{u}rich, 8093 Z\"{u}rich, Switzerland}
%ORCID 0000-0002-0026-743X

\author{L.~Facheris}
    \affiliation{Laboratory for Solid State Physics, ETH Z\"{u}rich, 8093 Z\"{u}rich, Switzerland}
%ORCID 0000-0001-8250-527X

\author{S.~Velja}
    \affiliation{Laboratory for Solid State Physics, ETH Z\"{u}rich, 8093 Z\"{u}rich, Switzerland}

\author{D.~Blosser}
    \affiliation{Laboratory for Solid State Physics, ETH Z\"{u}rich, 8093 Z\"{u}rich, Switzerland}
%ORCID 0000-0001-6653-3318

\author{Z.~Yan}
    \affiliation{Laboratory for Solid State Physics, ETH Z\"{u}rich, 8093 Z\"{u}rich, Switzerland}
%ORCID 0000-0002-9073-6358

\author{S.~Gvasaliya}
    \affiliation{Laboratory for Solid State Physics, ETH Z\"{u}rich, 8093 Z\"{u}rich, Switzerland}
%ORCID 0000-0001-7972-5909

\author{A.~Zheludev}
    \email{zhelud@ethz.ch}
   \homepage{http://www.neutron.ethz.ch/}
    \affiliation{Laboratory for Solid State Physics, ETH Z\"{u}rich, 8093 Z\"{u}rich, Switzerland}

\begin{abstract}
    We have found an unusual competition of two frustration mechanisms in the 2D quantum antiferromagnet  Cs$_2$CoBr$_4$. The key actors are the alternation of single-ion planar anisotropy direction of the individual magnetic Co$^{2+}$ ions, and their arrangement in a distorted triangular lattice structure.
    In particular, the uniquely oriented Ising-type anisotropy emerges from the competition of easy-plane ones, and for a magnetic field applied along this axis one finds a cascade of five ordered phases at low temperatures. Two of these phases feature magnetization plateaux. The low-field one is supposed to be a consequence of a collinear ground state stabilized by the anisotropy, while the other plateau bears characteristics of an ``up-up-down'' state exclusive for lattices with triangular exchange patterns. \textcolor[rgb]{0.00,0.00,0.00}{Such coexistence of the magnetization plateaux is a fingerprint of competition between the anisotropy and the geometric frustration in Cs$_2$CoBr$_4$}.
    \end{abstract}

\date{\today}
\maketitle

\section{Introduction}

A conventional picture of a frustrated quantum magnet implies a competition between the Heisenberg terms in a $S=1/2$ Hamiltonian. A Heisenberg magnet on a generic triangular lattice is an archetype example~\cite{Starykh_RepPrPhys_2015_TriangularReview}. Anisotropy, if present, is usually just a weak perturbation stemming from the spin-orbit interactions. Alternatively, like in a triangular lattice XXZ model, it acts in the same way on every bond and this situation is not drastically different from the Heisenberg case~\cite{Yamamoto_PRL_2014_TriangLatXXZ,*SellmannZhangEggert_PRB_2015_AnotherXXZtriangularPhD}. However, recently emerging topics of quantum spin ice~\cite{RossSavary_PRX_2011_QuantumIceNeutrons,*TaillefumierBenton_PRX_2017_QSicePhases,GingrasMcClarty_RepProgPhys_2014_QSIreview} or Kitaev magnets~\cite{Kitaev_AnnPhys_2006_Kitaev,WinterTsirlin_JPCM_2017_KitaevReview,*TakagiTakayama_NatRevPhys_2019_KitaevReview} teach us a very different approach. In those, anisotropy is the key player and the main ingredient creating frustration. Interestingly, this physics stemming from strong spin-orbit coupling is not endemic of the $4d$ and $4f$ magnets, but is also possible in $3d$ magnets, for instance cobalt-based ones~\cite{LiuChaloupkaKhaliullin_PRL_2020_CobaltKitaev}. In fact, in low-symmetry Co$^{2+}$ magnets ($S=3/2$ and quenched orbital momentum) the single-ion anisotropy that splits the $\ket{\pm1/2}$ and $\ket{\pm3/2}$ spin states may not be uniform between the sites. If no unique anisotropy axis is present, the interactions between the spins become frustrated automatically. If the spins are at the same time residing on a non-bipartite lattice such as a triangular one, geometrical frustration is also there. Two frustration mechanisms are present simultaneously and this results in a complicated interplay. This possibility is relatively well explored for a perfect triangular lattice~\cite{ZhuMaksimov_PRL_2018_AnisotropicTriangLattice,*MaksimovZhu_PRX_2019_AnisotropicTriangLattice}, but much less so for less symmetric cases.

The subject material of the present article, \CCoB, possesses an interesting combination of geometric frustration and anisotropy very much in line with the above discussion. It is the last unexplored member of the otherwise well known family of quantum magnets with the distorted triangular lattice Cs$_{2}MX_4$, where $M$ is copper or cobalt and $X$ is chlorine or bromine. The other three materials, essentially chain-like magnets \CCuC~\cite{Coldea_PRL_2001_CCC2D,*Tokiwa_PRB_2006_Cs2CuCl4phases,*SmirnovPovarov_PRB_2012_ESRordered,*SchulzeArsenijevic_PRR_2019_CCuC1Dheattrans,Starykh_PRB_2010_Cs2CuCl4theory}, \CCoC~\cite{Figgis_JChemSoc_1987_Cs2CoCl4refinement,KenzelmannColdea_PRB_2002_CsCoCldiffraction,BreunigGarst_PRL_2013_CsCoCl_TF,Breunig_PRB_2015_CsCoClPhD}, and more two-dimensional \CCuB~\cite{OnoTanaka_PRB_2003_CCBplateau,*TsujiiRotundu_PRB_2007_Cs2CuBr4UUD,FortuneHannahs_PRL_2009_CCBtransitions}, demonstrate very rich phase diagrams in applied magnetic fields. Although the existence of the last material in this quartet, \CCoB, was documented a long time ago~\cite{SeifertKhudair_JInorgNucChem_1975_A2CoX4bridgeman,*Seifert_ThAct_1977_A2BX4summary}, it was never investigated in the context of quantum magnetism. In this paper we report the highly unusual magnetic phase diagram of \CCoB, that is very anisotropic and features a cascade of magnetization plateaux for one particular direction of the magnetic field. One of these plateaux is found at zero magnetization, while the other corresponds to a field-induced ``up-up-down'' phase that is characteristic for the triangular lattice systems. The plateaux are very compatible with the effective Hamiltonian, which at the same time creates a lot of uncertainty for the nature of the remaining phases due to the unusual interplay of different frustration mechanisms.

\begin{figure}
  \centering
  \includegraphics[width=0.5\textwidth]{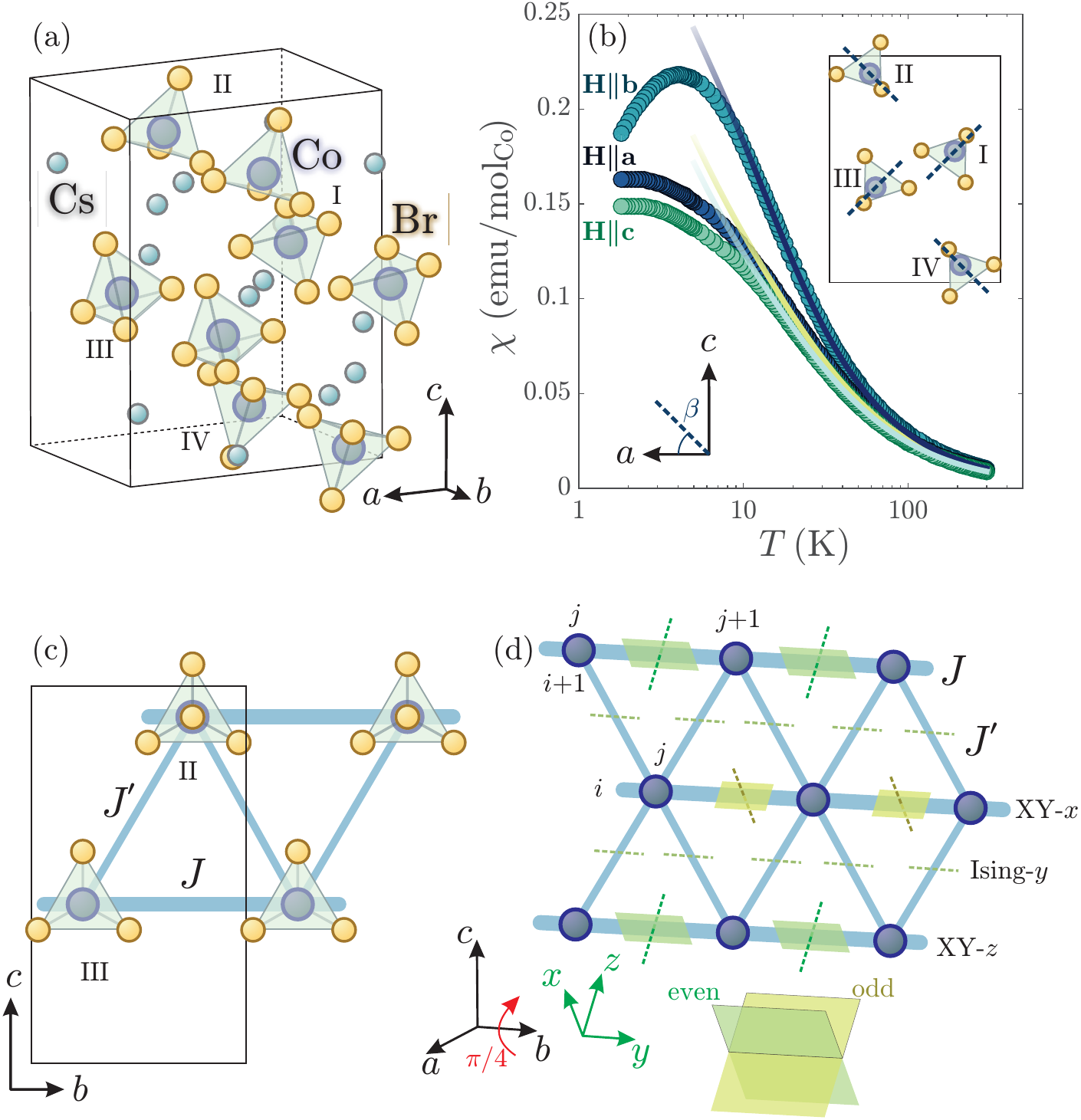}
  \caption{(a) Crystal structure of \CCoB. Four types of CoBr$_4$ tetrahedra are labeled as I-IV. (b) Magnetic susceptibility along three main crystallographic directions and the corresponding mean-field fit (solid lines; see text). Insets show the relative orientation of the anisotropy direction in different tetrahedra. (c) Distorted triangular lattice bond pattern in the $bc$ plane. (d) Effective anisotropies in the pseudospin-$1/2$ Hamiltonian. The $abc$ and $xyz$ coordinate systems are rotated by $\pi/4$ with respect to each other.}\label{FIG:XtalChi}
\end{figure}

\begin{figure}
  \centering
  \includegraphics[width=0.5\textwidth]{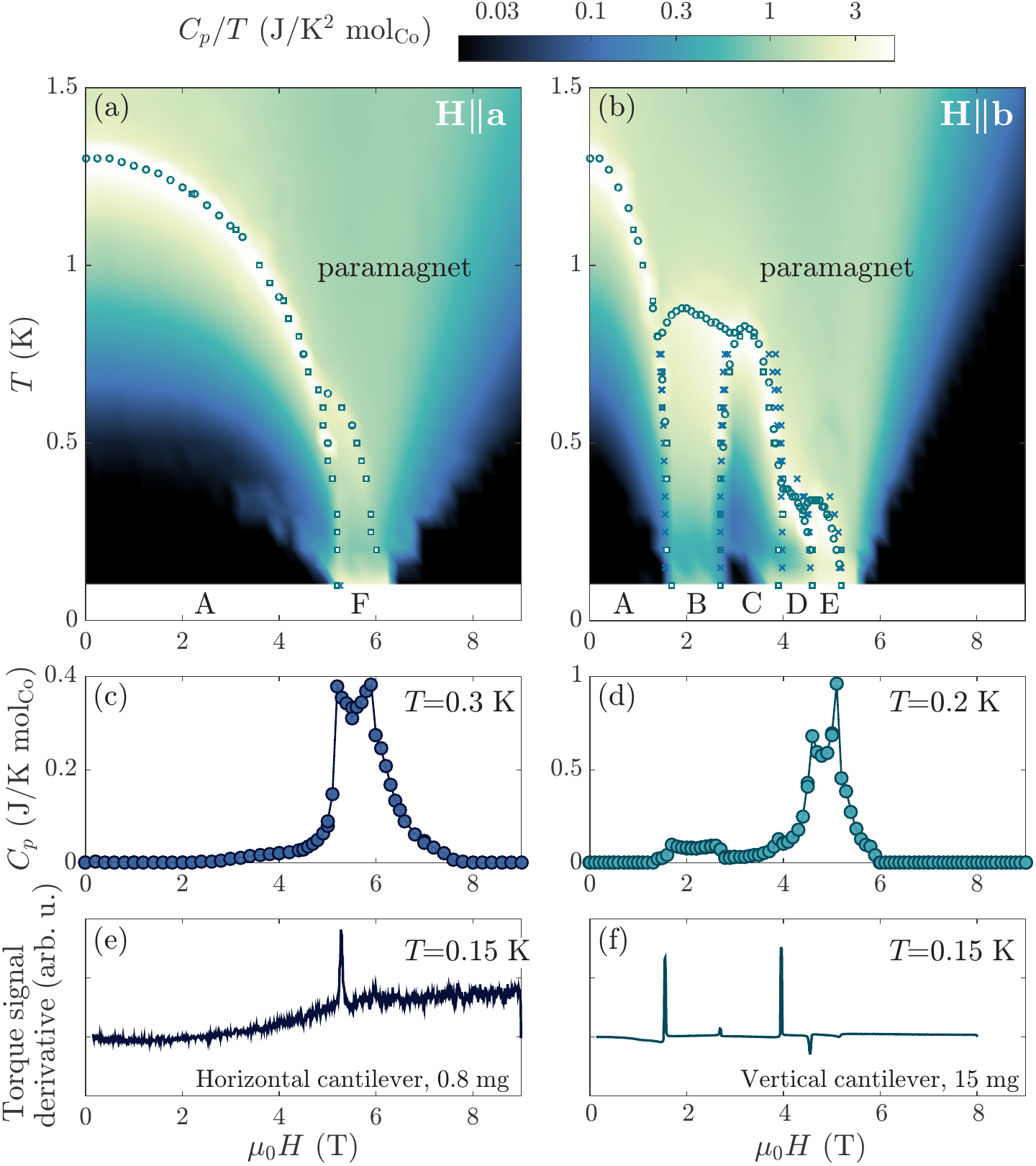}
  \caption{(a), (b) Specific heat for transverse $\mathbf{H}\parallel\mathbf{a}$ and longitudinal $\mathbf{H}\parallel\mathbf{b}$ applied magnetic fields. Circles identify the lambda anomalies in the $C_{p}(T)$ scans with fixed $H$, and squares --- in the $C_{p}(H)$ scans with fixed $T$. Crosses identify anomalies in the low temperature magnetometric measurements. (c), (d) Example $C_{p}(H)$ scans. (e), (f) Magnetic torque signal derivatives; geometry and approximate sample mass (resulting in a quite different sensitivity) are indicated on the plot. The original curves are displayed in Fig.~\ref{FIG:TRQs}.}\label{FIG:Cps}
\end{figure}

\section{Magnetic susceptibility and the effective spin Hamiltonian}
\subsection{Structural considerations}
Transparent, cerulean-colored single crystals of \CCoB\ were grown using the Bridgman method~\cite{SeifertKhudair_JInorgNucChem_1975_A2CoX4bridgeman}. Its structure is isomorphic to that of the other Cs$_{2}MX_4$ materials, orthorhombic $P\mathrm{nma}$ (space group $62$) with $a=10.193(2)$, $b=7.725(3)$, $c=13.510(4)$~\AA. The details of the sample growth and structure refinement are summarized in Appendix~\ref{APP:xray}. The unit cell shown in Fig~\ref{FIG:XtalChi}(a) contains four Co$^{2+}$ $S=3/2$ ions within four CoBr$_4$ distorted tetrahedra, related to each other by the mirror reflections in $ab$ and $bc$ planes. The mirror $ac$ plane is the only symmetry of an individual  distorted tetrahedron. As the local symmetry at the  Co$^{2+}$ site is lower than cubic, the single-ion anisotropy $D(\mathbf{n}\cdot\spop)^{2}$ should be present. The anisotropy axis is $\mathbf{n}=(\pm\cos\beta,0,\sin\beta)$ on the different tetrahedra, as the symmetry dictates. The angle $\beta$ and the sign of anisotropy constant $D$ are not known \textit{a priori} (in a sister material \CCoC\ they are estimated as $\beta\simeq51^{\circ}$ and $D\simeq7$~K~\cite{KenzelmannColdea_PRB_2002_CsCoCldiffraction}). Further ideas about the \emph{interactions} between the cobalt spins can be derived from comparison with \CCoC, \CCuC\ and \CCuB. All of them have the dominant interaction $J$ within the chains running along the $\mathbf{b}$ direction, while the weaker zigzag exchange $J'$ connects the chains into the distorted triangular lattice in the $bc$ plane as shown in Fig.~\ref{FIG:XtalChi}(c). The exchange along the $\mathbf{a}$ direction is negligibly small. The value of $J'/J$ may vary from almost zero (\CCoC\ case) to $0.3-0.5$ in the copper based members of the family.

\subsection{Magnetic susceptibility}
The key parameters of the Hamiltonian, such as $D$, $\beta$ and the mean-field exchange coupling $J_{0}=2J+4J'$ can be straightforwardly extracted from the magnetic susceptibility data for fields, applied along the three principal directions of the crystal.
Magnetic susceptibility $\chi=M/H$ of an $m=8.9$~mg \CCoB\ single crystal was measured with a MPMS SQUID magnetometer in a field of $0.1$~T. These data are shown in Fig.~\ref{FIG:XtalChi}(b). The $\mathbf{H}\parallel\mathbf{b}$ (perpendicular to $\mathbf{n}$) susceptibility is quite different from the susceptibilities for $\mathbf{H}\parallel\mathbf{a,c}$ directions that look rather similar (angles $\beta$ and $\pi/2-\beta$ between the field and $\mathbf{n}$). All of them show the typical ``Curie tail'' behavior at high temperatures, that becomes suppressed at low temperatures as the antiferromagnetic correlations take over. Susceptibility along the $\mathbf{b}$ direction shows a rounded maximum close to $4$~K --- a picture typical for the low-dimensional magnets with suppressed magnetic order. No signs of ordering are found down to $1.8$~K. The simultaneous fitting of the data for all three directions, based on a single-ion model with the mean-field interactions (the details of the fit are given in the Appendix~\ref{APP:Fit}), yields $D=14(1)$~K, $\beta=44(1)^{\circ}$, and $J_{0}=5.5(2)$~K, with the $g$ factors being $2.42(1),~2.47(2)$ and $2.37(1)$ along the $\mathbf{a},\mathbf{b}$ and $\mathbf{c}$ directions. This means that (i) the single-ion anisotropy is of an \emph{easy-plane} type, so at the low temperature only the pseudospin-$1/2$ degrees of freedom are active, and (ii) the easy planes, while being uniform within the chains, have alternating orientation between the chains and the neighboring ones are \emph{nearly orthogonal} to each other. We can consider $\beta=\pi/4$ for practical purposes. Then, utilizing the ``rotated'' $xyz$ coordinate system [Fig.~\ref{FIG:XtalChi}(d)], the approximate Hamiltonian for the $S=3/2$ cobalt spins can be written as

\begin{align}\label{EQ:Hamilt}
  \hamilt_{3/2}=&\sum_{i,j}D\left[\left(\spin_{2i,j}^{z}\right)^{2}+\left(\spin_{2i+1,j}^{x}\right)^{2}\right]+J(\spop_{i,j}\cdot\spop_{i,j+1})\\ \nonumber
  &+J'(\spop_{i,j}\cdot\spop_{i+1,j})+J'(\spop_{i,j}\cdot\spop_{i+1,j+1}).
\end{align}

\subsection{Effective Hamiltonian}

To construct the effective low-energy Hamiltonian, one needs to project out the high-spin states that are inaccessible at low temperatures due to large $D$. This is achieved by the Schrieffer–-Wolff transformation~\cite{SchriefferWolff_PR_1966_SchriefferWolff,*Bravyi_AnnPhys_2011_SchriefferWolffReview,BreunigGarst_PRL_2013_CsCoCl_TF}, where to the zeroth order we can simply replace the spin-$3/2$ operators with the spin-$1/2$ ones as $\spin^{x,y}\rightarrow2\spin^{x,y}$ and $\spin^{z}\rightarrow\spin^{z}$ in the even chains, and $\spin^{z,y}\rightarrow2\spin^{z,y}$ and $\spin^{x}\rightarrow\spin^{x}$ in the odd chains. The resulting Hamiltonian is

\begin{align}\label{EQ:XYZHamilt}\nonumber
  \hamilt_{1/2}=&\sum_{i,j}4J(\spop_{i,j}\cdot\spop_{i,j+1})\\\nonumber
  -3J&\left[\spin_{2i,j}^{z}\spin_{2i,j+1}^{z}+\spin_{2i+1,j}^{x}\spin_{2i+1,j+1}^{x}\right]\\
  &+2J'(\spop_{i,j}\cdot\spop_{i+1,j})+2J'(\spop_{i,j}\cdot\spop_{i+1,j+1})\\
  &~+2J'\spin_{i,j}^{y}\spin_{i+1,j}^{y}+2J'\spin_{i,j}^{y}\spin_{i+1,j+1}^{y}.\nonumber
\end{align}

\begin{figure}
  \centering
  \includegraphics[width=0.5\textwidth]{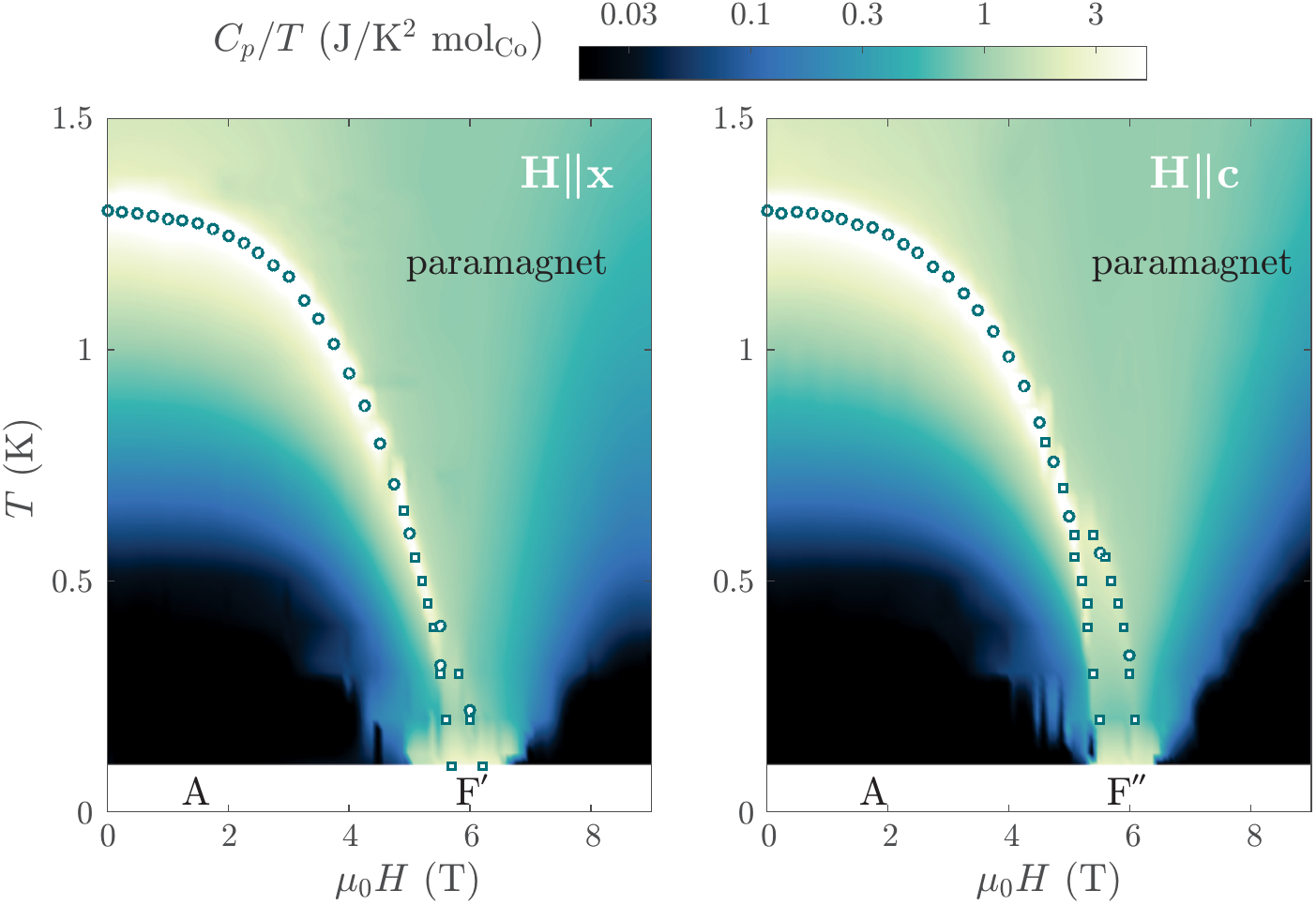}
  \caption{Phase diagrams for the $\mathbf{H}\perp \mathbf{b}$ field directions. Color shows $C_{p}/T$ data; points (circles and squares) are the lambda peak positions identified in the temperature or field scans correspondingly.}\label{FIG:AxCdiagramsCP}
\end{figure}

\begin{figure}
  \centering
  \includegraphics[width=0.5\textwidth]{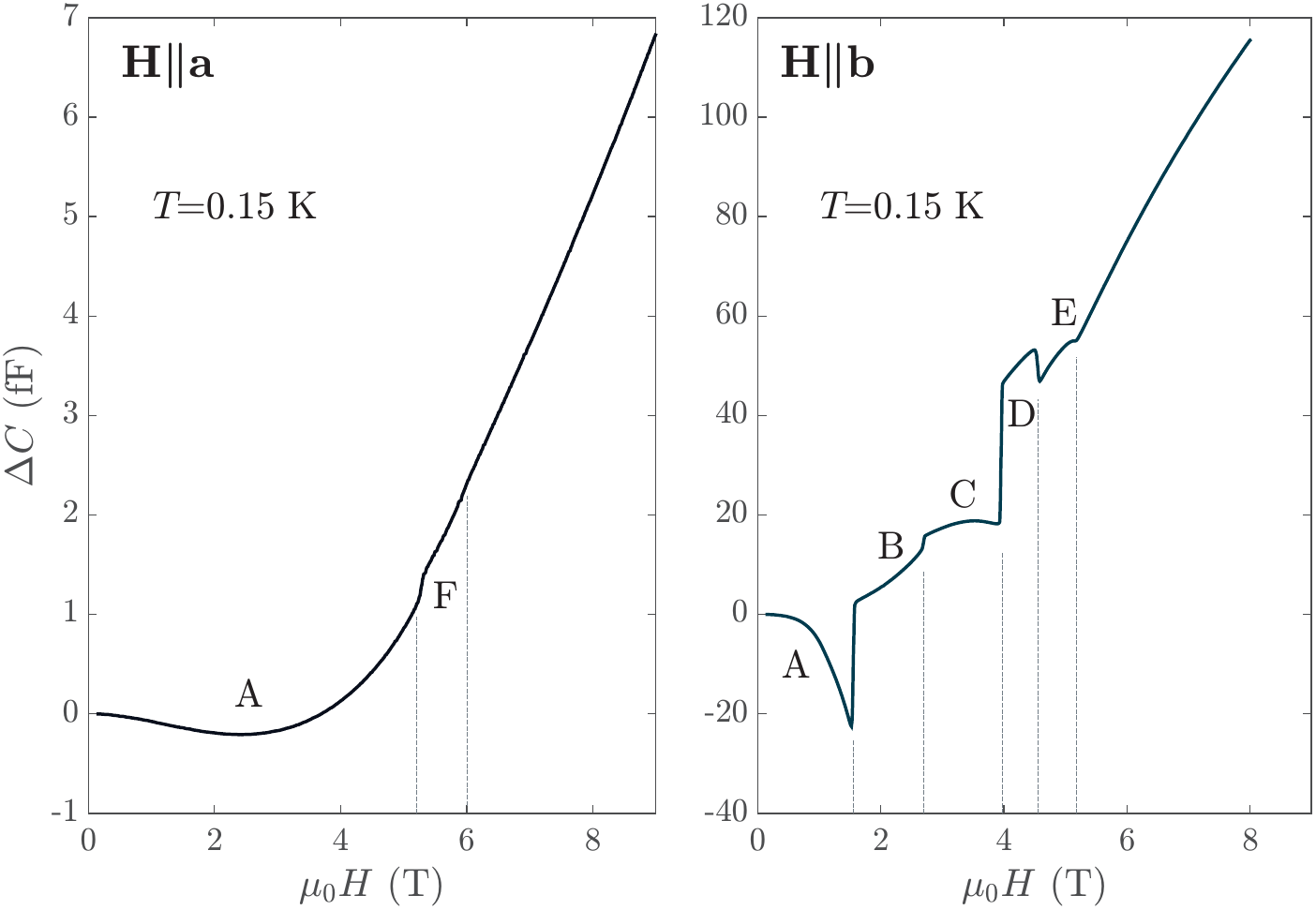}
  \caption{Magnetic torque data for $\mathbf{H}\parallel\mathbf{a}$ and  $\mathbf{H}\parallel\mathbf{b}$ direction at $0.15$~K. A cantilever-to-base electric capacitance change is shown as the function of field, that is increasing at $10^{-3}$~T/s. The corresponding derivatives are displayed in Fig.~\ref{FIG:Cps}. }\label{FIG:TRQs}
\end{figure}

\begin{figure}
  \centering
  \includegraphics[width=0.5\textwidth]{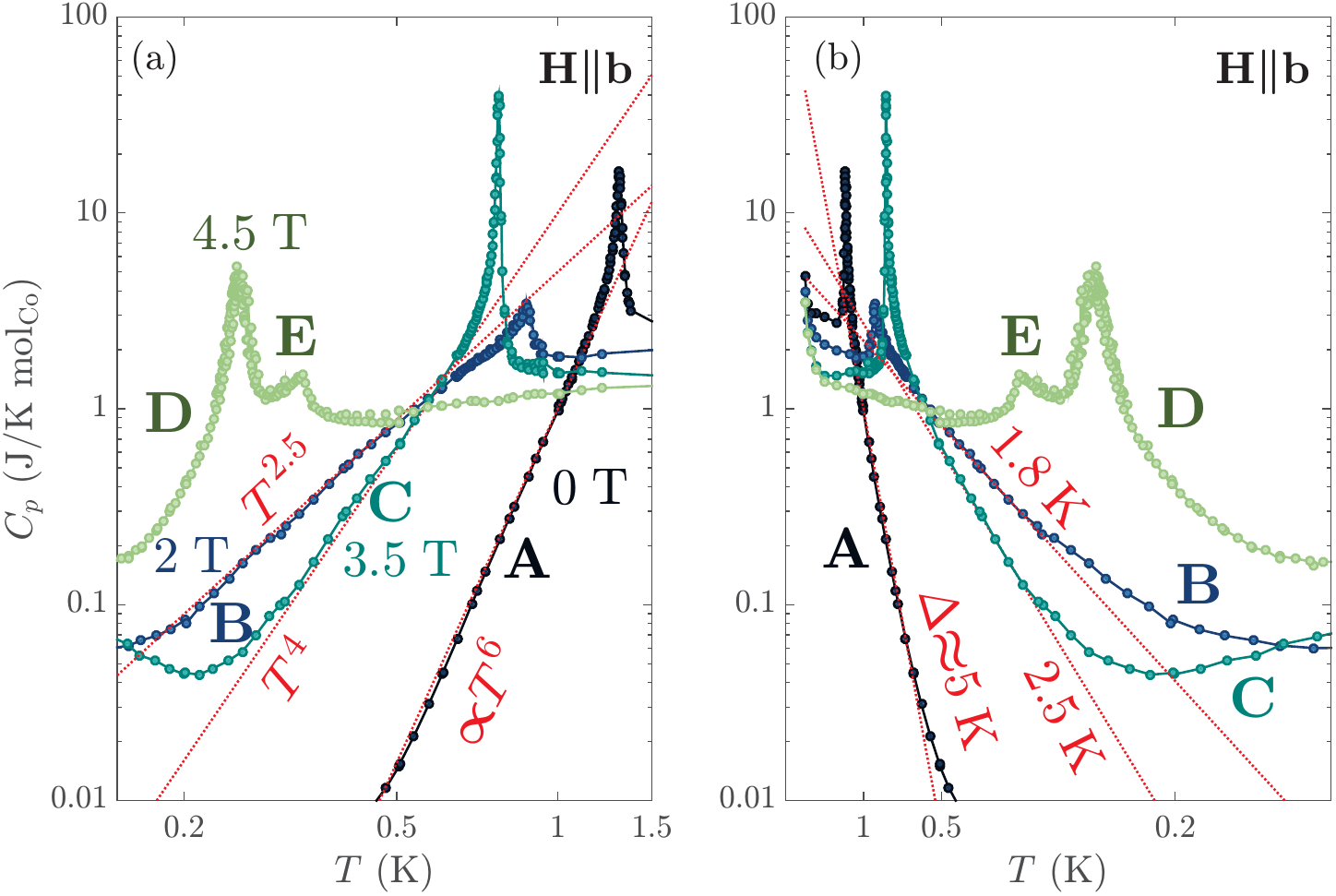}
  \caption{Specific heat at several $\mathbf{H}\parallel\mathbf{b}$ magnetic fields. Panel (a) is the logarithmic, and panel (b) is the Arrhenius plot of the same data.
  The approximate key parameters describing the linearized regime (dashed lines) according to Eq.~(\ref{EQ:powerlaw}) in (a) or to Eq.~(\ref{EQ:gappedlaw}) in (b) are indicated in the plots. The low-temperature upturns are probably the nuclear specific heat coming into play. }\label{FIG:BTdeps}
\end{figure}

\begin{figure}
  \centering
  \includegraphics[width=0.5\textwidth]{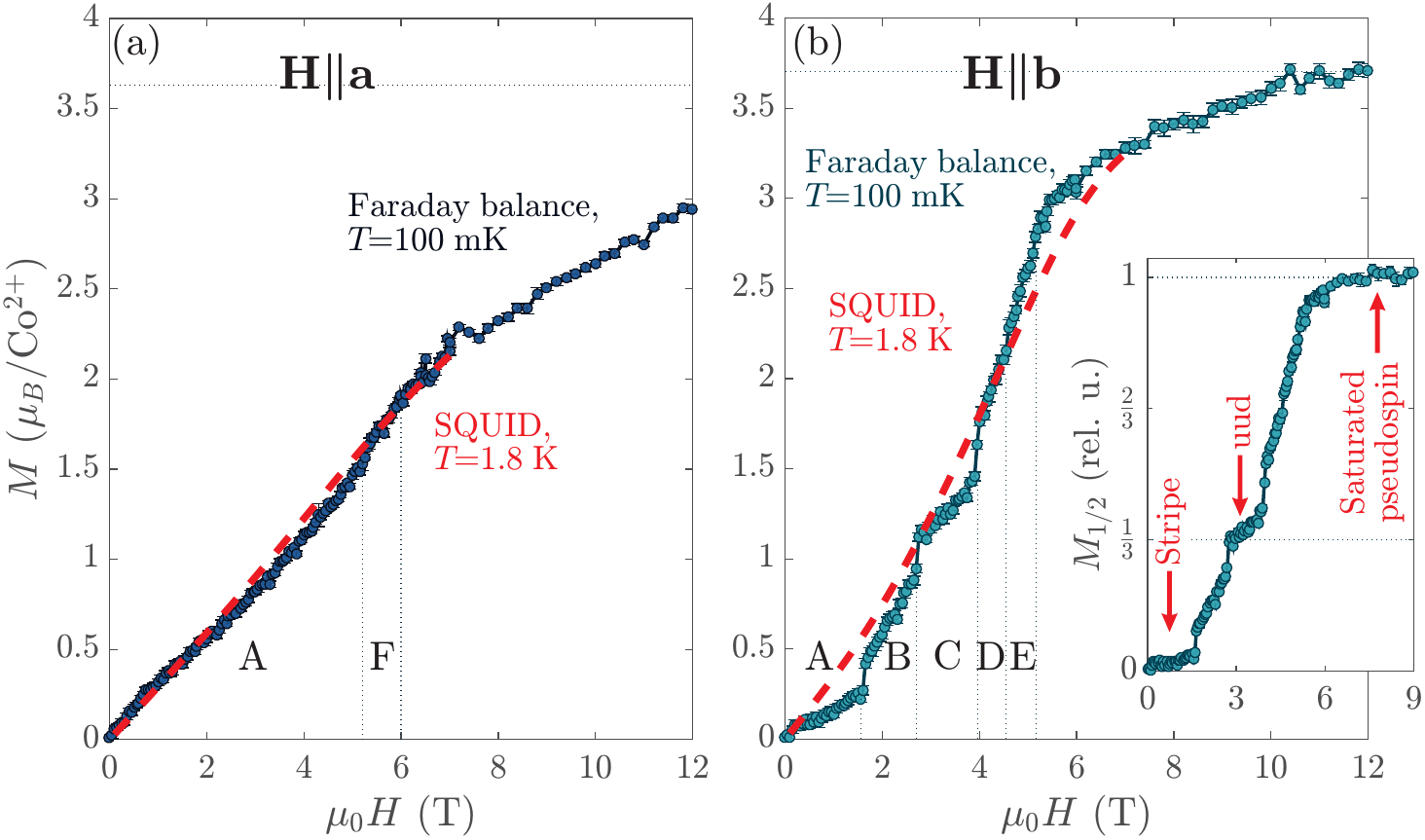}
  \caption{(a,b) Magnetization of \CCoB\ at $T=0.1$~K measured with the Faraday balance technique. Thick dashed line is the reference SQUID data at $1.8$~K [same experiment as in Fig.~\ref{FIG:XtalChi}(b)]. Horizontal dashed line shows the expected saturation moment $g\mu_{B}S$. Inset in (b) shows the relative magnetization of the pseudospin-$1/2$ degrees of freedom derived from the magnetization curve. Three plateaux are identified as corresponding to the collinear antiferromagnetic, collinear up-up-down, and fully polarized arrangements of the pseudospins.}\label{FIG:M}
\end{figure}

A graphical representation of this Hamiltonian is also given in Fig.~\ref{FIG:XtalChi}(d). Here the intrachain exchange is of a strongly XY nature, with the easy-plane direction alternating between the chains. In contrast, the frustrated interchain interaction is now Ising-like, with the easy axis given by the only common direction of two adjacent easy planes. This special direction coincides with the structural $\mathbf{b}$ axis.

\section{Specific heat and the phase diagrams}
\subsection{Transverse direction}

The difference between the emergent Ising axis $\mathbf{b}$ and the other directions is clearly manifest in the specific heat data. Measurements on an $m=0.81(6)$~mg \CCoB\ sample were carried out on a $9$~T PPMS system with a $^3$He-$^4$He dilution refrigerator insert. A standard relaxation calorimetry method was used, also in combination with the so-called ``long-pulse'' technique~\cite{Scheie_JLTP_2018_LongHeatPulse}. The resulting cumulative specific heat data set for $\mathbf{H}\parallel\mathbf{a,b}$ directions is shown in Fig.~\ref{FIG:Cps}(a,b). For the transverse field direction $\mathbf{a}$ the phase diagram essentially contains a single ordered ``A'' phase and a small ``F'' satellite.

An almost identical picture is observed for the other $\mathbf{H}\perp\mathbf{b}$ directions as summarized in Fig.~\ref{FIG:AxCdiagramsCP}. Like with the magnetic susceptibility, the $\mathbf{a}$ and $\mathbf{c}$ directions look very similar. An additional phase diagram is measured in a field orientation $\mathbf{H}\parallel \mathbf{x}$, with the $\mathbf{x}$ direction being crudely at $45^{\circ}$ between $\mathbf{a}$ and $\mathbf{c}$. The F phase is almost suppressed here and the phase boundary becomes nearly linear. The possible misalignment from the intended direction in this measurement can be estimated as $10^{\circ}-15^{\circ}$. Thus, it may be possible that the F-phase completely disappears at some close field orientation.

\subsection{Longitudinal direction}

In contrast, the magnetic field along the emergent Ising axis $\mathbf{b}$ results in a sequence of \emph{five} different phases, from ``A'' to ``E''. In either case the phase diagram is terminated around $5.5-6$~T and above this field the system is simply a semipolarized anisotropic paramagnet. This is in line with the effective exchange coupling $J_{0}\simeq5.5$~K determined from the mean-field analysis. The anomalies corresponding to the phase transitions are quite visible in the $C_{p}(H)$ scans, with the examples given in Fig.~\ref{FIG:Cps}(c,d). Additional insight is brought by the capacitive torque magnetometry. Similarly to~\cite{FengPovarov_PRB_2018_LinariteTilted}, the sample is placed on a flexible cantilever and the force that it experiences is measured by the cantilever deflection that translates into the setup's capacity change. More details on the technique are given in Appendix~\ref{APP:torque}. The torque data in Figs.~\ref{FIG:Cps}(e,f) and~\ref{FIG:TRQs} ensure that all the observed transitions are of magnetic origin: each transition results in a strong anomaly in the total force that acts on the sample in the magnetic field.

In Fig.~\ref{FIG:BTdeps} one can see a comparative plot of $C_{p}(T)$ dependencies for $\mathbf{H}\parallel\mathbf{b}$ in the A, B, C and D/E phases. Panel (a) is showing the logarithmic plot of the data, assuming its interpretation as a power law:

\begin{equation}\label{EQ:powerlaw}
  C_{p}(T)=\mathcal{A}T^{\alpha}.
\end{equation}

Panel (b) demonstrates the same data, but in a $\log{C_{p}}$ vs $T^{-1}$ representation (Arrhenius plot). This assumes a thermally activated dependence:

\begin{equation}\label{EQ:gappedlaw}
  C_{p}(T)=\mathcal{\tilde{A}}e^{-\Delta/T}.
\end{equation}

In Fig.~{\ref{FIG:BTdeps}} one can compare how the two possible descriptions fit to the data. According to Eq.~(\ref{EQ:powerlaw}) one finds $T^6$ and $T^4$ in the A and C phases correspondingly ($0$ and $3.5$~T data sets). Such strong power laws cast some doubt on Eq.~(\ref{EQ:powerlaw}) correctly reflecting the physics of the problem. In contrast, thermally activated description~(\ref{EQ:gappedlaw}) allows us to estimate the corresponding gaps as $5$ and $2.5$~K, which is in principle consistent with the Hamiltonian parameters. On the other hand, for the B phase ($2$~T data set) the power law $T^{2.5}$ seems to provide a reasonable description of the data. It is also in line with the simple $T^{d/z}$ expectation for a gapless material with a linear $z=1$ dispersion   and  an effective dimensionality being in between $d=2$ and $d=3$. The suppressed magnetic susceptibility in the A and C phases, and the substantial susceptibility in the B phase are also consistent with the understanding of these phases as the gapped and gapless ones.

The last scan at $4.5$~T traverses both the D and E phases. While the critical behavior obscures the ``true'' temperature dependence of the magnetic specific heat, it is interesting to note that above the transition $C_p$ is almost temperature independent. Together with the reduced ordering temperatures, this marks the E and D phases as some ``satellite'' states emerging in the vicinity of the quantum critical point due to the competition between the small parameters. The same statement is equally applicable to the F-phases in the transverse field orientation. The principal phases A, B, and  C seem to be much more robust and of a different nature.

\section{Magnetization plateaux}

While for small magnetic fields the gap seems a natural consequence of the Ising-like anisotropy, its presence in the magnetized C state is not so trivial. A candidate gapped state in a system featuring a triangular bond pattern is the famous ``up-up-down'' (abbreviated as uud) collinear spin arrangement. This is further confirmed by a direct measurement of the \CCoB\ magnetization curve at $100$~mK. This measurement is performed on the same sample as the specific heat with the help of a miniature home-built Faraday balance magnetometer with a twist-resistant cantilever~\cite{Blosser_RevSciInstr_2020_FaradayBalance}. The resulting curves are demonstrated in Fig.~\ref{FIG:M} together with the reference data from a SQUID magnetometer at $1.8$~K that was also used for the calibration. While for the transverse $\mathbf{H}\parallel \mathbf{a}$ direction the measured magnetization curve is relatively smooth and shows only the weak kinks at the two phase transitions, the situation is very different for the longitudinal $\mathbf{H}\parallel \mathbf{b}$ magnetization. Most of the transitions are marked with discontinuities. Moreover, the slope of the magnetization curve is clearly reduced in the A and C phases. But are these the real magnetization plateaux? We argue that they are. The extra slope $dM/dH$ is originating from admixing of the single-ion high-energy $\ket{\pm3/2}$ states to the ground state by a noncommuting magnetic field, and it is also pronounced at high fields when the pseudospin degrees of freedom are fully polarized. The slope of the magnetization curve slightly above $6$~T  should provide a reasonable estimate of the effect (at high magnetic field this effect is reduced, as the magnetization curve flattens in general). The corrected data representing the relative magnetization of the pure pseudospin-$1/2$ are shown in the inset of Fig.~\ref{FIG:M}(b). The plateau character of the A and C phases is \textcolor[rgb]{0.00,0.00,0.00}{well} pronounced in this representation.

\section{Discussion}
\subsection{Possible collinear structures}

For the low-field A phase a suitable candidate structure might be a collinear antiferromagnetic ``stripe'' state (as in a sister material \CCoC~\cite{KenzelmannColdea_PRB_2002_CsCoCldiffraction}; see Fig.~\ref{FIG:States}). This state automatically satisfies the anisotropic exchange interactions within both even and odd chains. On the mean-field level the chains remain decoupled for any strength of $J'$, and the overall collinear structure must be fixed by some kind of ``order from disorder'' mechanism~\cite{Henley_PRL_1989_OrderByDisorder}. This state is significantly more robust in \CCoB\ than in \CCoC: it is destabilized by a field strength that is nearly $0.3$ of the saturation field for the $1/2$-pseudospins, while in the latter material the corresponding number is only $0.1$. This may reflect the increased $J'/J$ ratio in \CCoB. To summarize, the A plateau is naturally explained as the nonmagnetized state of a collinear magnet in a small field applied along the effective easy axis.

 The magnetization C plateau is close to $1/3$ of the full saturated value for the pseudospin, validating it as a collinear uud structure. This is again pointing to the importance of $J'$ bonds. The uud collinear structures are specific to the systems with triangular exchange patterns. Being stabilized by the quantum fluctuations at low temperatures, they represent another example of ``order from disorder''~\cite{ChubukovGolosov_JPCM_1991_QuantumUUD,Starykh_RepPrPhys_2015_TriangularReview}. Again, the effectively easy-axis character of the system will be in favor of such structure too. The anisotropy and the quantum fluctuations are \textcolor[rgb]{0,0,0}{acting together}, and the resulting $1/3$ magnetization plateau is \textcolor[rgb]{0,0,0}{enormously} wide: it occupies almost $0.25$ of the full phase diagram width in the magnetic field. For comparison, in \CCuB\ the relative width of the uud phase is just $0.05$~\cite{OnoTanaka_PRB_2003_CCBplateau,*TsujiiRotundu_PRB_2007_Cs2CuBr4UUD,FortuneHannahs_PRL_2009_CCBtransitions}, and in the ideal triangular Heisenberg case the expected number is $0.2$~\cite{ChubukovGolosov_JPCM_1991_QuantumUUD}. Remarkably, a group of recently reported delafossite-like anisotropic triangular lattice antiferromagnets, such as NaYbO$_2$~\cite{RanjithDmytriieva_PRB_2019_NaYbO2first,*DingManuel_PRB_2019_NaYbO2gaplessproposal,*Bordelon_NatPhys_2019_NaYbO2uud}, NaYbSe$_2$~\cite{RanjithLuther_PRB_2019_NaYbSe2uud} and NaYbS$_2$~\cite{BaenitzSchlender_PRB_2018_NaYbS2first,*MaLiGao_arXiv_2020_NaYbS2uud}, were found to exhibit an unusually wide uud phase as well. A property they share with \CCoB\ is the significant anisotropy that varies between the bonds. \textcolor[rgb]{0.00,0.00,0.00}{However, the simultaneous presence of another wide plateau at $M=0$ is a special feature of \CCoB.}

\subsection{\CCoB\ and the close materials}

\begin{figure}
  \centering
  \includegraphics[width=0.5\textwidth]{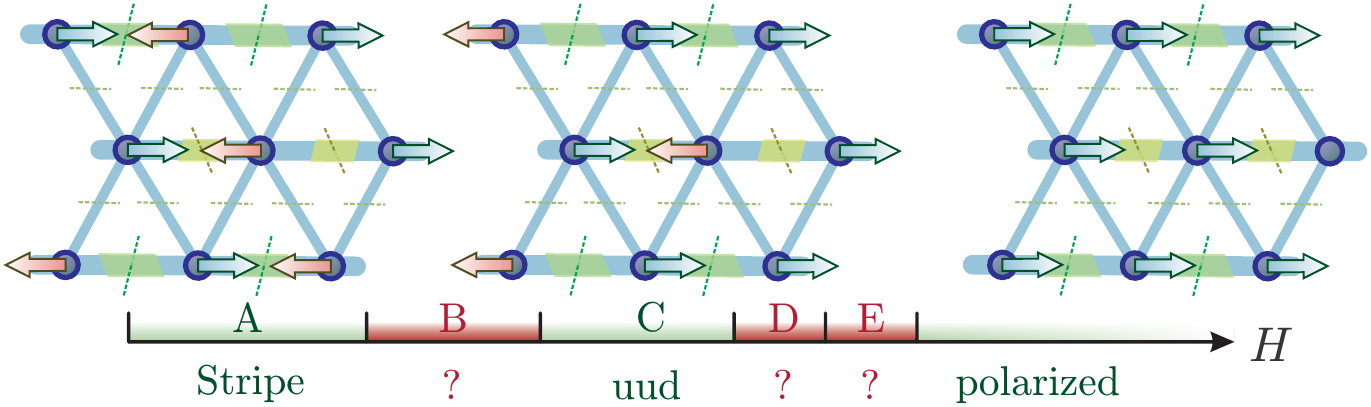}
  \caption{Sketches of the plausible collinear magnetic phases (stripe, uud, and saturated) in \CCoB\ for the $\mathbf{H}\parallel\mathbf{b}$ field direction. The intervening phases B, D, and E remain to be clarified.}\label{FIG:States}
\end{figure}

The nature of the remaining phases, B,D,E, and F, is unknown at the moment. While in the XXZ-type models or in the presence of weak spin-orbit interactions various (nearly) coplanar phases are known to occur in a magnetized triangular lattice~\cite{ChubukovGolosov_JPCM_1991_QuantumUUD,ChenJuJiang_PRB_2013_DistortTriangquasi1D,GrisetHead_PRB_2011_DefTriangwithDM,Yamamoto_PRL_2014_TriangLatXXZ,*SellmannZhangEggert_PRB_2015_AnotherXXZtriangularPhD}, heavy frustration created by the competing single-ion anisotropy directions would probably be prohibitive for their formation in \CCoB. Spins in the neighboring chains strongly prefer to be confined in two orthogonal planes, and, while allowing the collinear states (as shown in Fig.~\ref{FIG:States}), this circumstance impedes the coplanar ones. \textcolor[rgb]{0.00,0.00,0.00}{The behavior of \CCoB\ is clearly different from a more conventional XXZ triangular-lattice magnets such as Ba$_3$CoSb$_2$O$_9$~\cite{Quirion_PRB_2015_Ba3CoSb2O9ultrasonicUUD,*KoutroulakisZhou_PRB_2015_Ba3CoSb2O9NMRUUD,KamiyaGe_NatComm_2018_Ba3CoSb2O9spectrumUUD}.}

The phase diagram of the sister material \CCoC, demonstrating some incommensurate and multi-$\mathbf{Q}$ states~\cite{KenzelmannColdea_PRB_2002_CsCoCldiffraction,Gosuly_2016_PhDthesis}, is of limited guidance too. It misses the aspect of significant frustration by $J'$ interactions, as it can be concluded from the absence of the uud state. Thus, neither the conventional triangular lattice nor the chain-based approach seem to be fully appropriate for the discussion of the \CCoB\ phase diagram. The situation that we encounter here according to Eq.~(\ref{EQ:XYZHamilt}) is more akin, although not fully identical, to a triangular Kitaev--Heisenberg model that can host  \textcolor[rgb]{0,0,0}{multiple exotic spin states}~\cite{BeckerHermanns_PRB_2015_TriangularKitaevGS,*RousochatzakisRossler_PRB_2016_TriangularKitaevGS,*KishimotoMorita_PRB_2018_HoneyTriangularKitaevGS}. Although the proposal for the extremely exotic physics is too preliminary at the moment, Hamiltonian~(\ref{EQ:XYZHamilt}) taken together with the phase diagram in Fig.~\ref{FIG:Cps}(b) is suggestive of some nontrivial spin textures that may be present among the many magnetic phases. We would also like to stress that the proposed Hamiltonian~(\ref{EQ:XYZHamilt}) is the most basic one, and does not include the further symmetry-allowed terms such as the second single-ion anisotropy constant $E$ and the multiple Dzyaloshinskii--Moriya interactions (that are very important in \CCuC, for instance~\cite{Starykh_PRB_2010_Cs2CuCl4theory,Coldea_PRL_2002_Cs2CuCl4highfield,*PovarovSmirnov_PRL_2011_ESRdoublet}). \textcolor[rgb]{0,0,0}{From the experimental point of view, the ``frustration ratio'' of $J_{0}/T_{N}\simeq6$ in \CCoB\ is quite the same as in the strongly anisotropic and frustrated magnet $\alpha-$RuCl$_3$~\cite{SearsSongvilay_PRB_2015_alphaRuClsusceptible,WinterTsirlin_JPCM_2017_KitaevReview,*TakagiTakayama_NatRevPhys_2019_KitaevReview}.}

The plateaux coexistence is the direct evidence of an interplay between the frustrated exchange and the anisotropies. \textcolor[rgb]{0.00,0.00,0.00}{There exists another material that shows the $M=0$ and $M=1/3$ plateaux simultaneously --- an Ising chain $\alpha$-CoV$_2$O$_6$~\cite{HeYamaura_JACS_2009_CoV2O6plateus}. In this compound the Co$^{2+}$ ions display a strong Ising anisotropy with a uniquely oriented axis. These ions form the ferromagnetic chains acting as the Ising ``superspins'',  transversely coupled in an antiferromagnetic triangular lattice way. Thus, the basic physics of $\alpha$-CoV$_2$O$_6$ can be described by the classical triangular lattice Ising model, featuring only $M=0,~1/3$ and $1$ magnetization states. This is what is observed in this material indeed~\cite{LenertzAlaria_PRB_2012_CoV2O6groundstates,*Markkula_PRB_2012_CoV2O6groundstates2,*KimKimKim_PRB_2012_CoV2O6groundstates3,SaulVodenicarevic_PRB_2013_CoV2O6theory} (although a closer investigation also reveals some extra metastable states at the abrupt magnetization steps~\cite{EdwardsLane_PRB_2020_CoV2O6metastable}). In contrast, the Ising-like character of \CCoB\ is not inherited from the uniaxial ionic anisotropy, but emerges from the competition between the ionic planar anisotropies on different sites. This circumstance, together with the triangular-like geometry, leads to a much richer phase diagram. Nonetheless, the Ising toy model illustrating the case of $\alpha$-CoV$_2$O$_6$ and discussed in more details in Appendix~\ref{APP:toy} provides a way for estimating the exchange ratio $J'/J$. From comparing the energies of stripe, uud, and polarized states as a function of this ratio and the magnetic field, we can crudely estimate $J'/J\sim0.3$ in \CCoB.}

\section{Conclusion}

To summarize, the $S=3/2$ quantum antiferromagnet \CCoB\ is found to feature an unusual type of frustration that stems from \emph{both} the geometry of the exchange bonds and the geometry of the strong single-ion anisotropies. The ``spin space'' component of the frustration creates an effective $S=1/2$ Hamiltonian with the bond-dependent exchanges. Coexistence of $M\simeq 0$ and $M\simeq 1/3$ magnetization plateaux is the \textcolor[rgb]{0,0,0}{exceptional} feature of \CCoB\ and it is the direct consequence of interplay between the anisotropy and exchange geometries. While the plateau states can be preliminarily identified as the collinear antiferromagnetic and uud structures naturally compatible with the effective Hamiltonian, the situation is much less certain for the magnetizable phases. \textcolor[rgb]{0,0,0}{Scenarios derived from the known cases of the XXZ-like triangular lattice or XY-like chains are equally problematic here.} To the best of our knowledge, the frustrated Hamiltonians of this type were not considered in the literature before. At the same time, the prototype material is already there and the corresponding parameters can easily be tuned by the chemical composition or the pressure. We believe that further experiments \textcolor[rgb]{0,0,0}{(neutron spectroscopy in particular)} and theoretical effort aimed at exploring this specific frustration mechanism may yield some novel exotic magnetic states.

\acknowledgments

This work was supported by Swiss National Science Foundation,
Division II. We would like to thank Dr. George Jackeli (University of Stuttgart) for illuminating discussions.

\appendix

\section{Crystal growth and crystal structure}
\label{APP:xray}

The crystal of \CCoB\ used in the present refinement procedure was grown from a stoichiometric $2:1$ mixture of CsBr (Sigma Aldrich, 99.9\%) and CoBr$_2$ (Sigma Aldrich, 99.99\%, anhydrous). The powders were mixed together and finely ground in an argon-filled glove box, then loaded in a glassy carbon crucible. The crucible was kept under high vacuum at $\simeq200^{\circ}$~C for three days, then sealed.  The Bridgman furnace growth protocol consisted of slow translation ($1.5$~cm/day, $10$~cm total) of the crucible through the point with a temperature of $570^{\circ}$~C and a gradient of about $10^{\circ}$~C/cm, followed by a slow cool-down.

The x-ray refinement of the \CCoB\ single crystal was performed with a Bruker APEX-II diffractometer at room temperature using 69072 reflections of which 1428 were unique, with the final $R$ factors $R=0.033$ and $wR=0.1264$. The results are given in Table~\ref{TAB:xray}.

\begin{table*}
  \centering
  \begin{tabular}{ l c c c c }
  \hline\hline
  Lattice parameters & $a$ & $b$& $c$\\
  \hline
  &  10.1931(19) & 7.725(3) & 13.510(4)\\
  & & & \\
    \hline
        Symmetry transformations & & & $P\mathrm{nma}$ \\
    \hline
        & $x$ & $y$ & $z$ \\
    & $\frac{1}{2}-x$ & $-y$ & $\frac{1}{2}+z$ \\
    & $-x$ & $\frac{1}{2}+y$ & $-z$ \\
    & $\frac{1}{2}+x$ & $\frac{1}{2}-y$ & $\frac{1}{2}-z$ \\
        & $-x$ & $-y$ & $-z$ \\
        & $-\frac{1}{2}+x$ & $y$ & $-\frac{1}{2}-z$ \\
        & $x$ & $-\frac{1}{2}-y$ & $z$ \\
        & $-\frac{1}{2}-x$ & $-\frac{1}{2}+y$ & $-\frac{1}{2}+z$ \\
    & & & \\
    \hline
      Atom & $x$ & $y$ & $z$ & Equiv. U$_{\text{iso}}$\\
      \hline
    Cs 1 & 0.52242(6) & 0.7500 & 0.32846(4) & 0.0378(2) \\
    Cs 2 & 0.86229(7) & 0.7500 & 0.60296(7) & 0.0577(3) \\
    Co 1 & 0.26443(10) & 0.7500 & 0.57771(8) & 0.0284(3) \\
    Br 1 & 0.49803(9) & 0.7500 & 0.59887(8) & 0.0498(3) \\
    Br 2 & 0.18717(10) & 0.7500 & 0.41040(7) & 0.0495(3) \\
    Br 3 & 0.17424(7) & 0.49823(9) & 0.65413(7) & 0.0564(3) \\
   \hline\hline
  \end{tabular}
  \caption{Chemical structure of \CCoB. The resulting relative atomic positions are rather close to ones in \CCoC~\cite{Figgis_JChemSoc_1987_Cs2CoCl4refinement} (the coordinate system differs by a $[\frac{1}{2},~\frac{1}{2},~\frac{1}{2}]$ offset in this reference).}\label{TAB:xray}
\end{table*}

\section{Susceptibility fitting procedure}
\label{APP:Fit}

We start from a single $S=3/2$ ion model with uniaxial anisotropy:

\begin{equation}\label{EQ:HamiltSI}
  \hamilt^{\mathrm{SI}}=D(\spop\cdot\mathbf{n})^{2}.
\end{equation}

  The anisotropy axis $\mathbf{n}$ is perpendicular to the $\mathbf{b}$ direction. Thus, for all the ions $\mathbf{H}\parallel \mathbf{b}$ is a purely transverse orientation. Susceptibility per ion is described by the formula

\begin{equation}\label{EQ:Chiperp}
  \chi_{\mathbf{b}}^{\mathrm{SI}}(T)=\frac{(g_{\mathbf{b}}\mu_{\mathrm{B}})^{2}}{T}\left[\dfrac{1}{1-\exp{\left(-2D/T\right)}}+\frac{3T}{4D}\tanh{\frac{D}{T}}   \right].
\end{equation}

The situation is more complicated for the remaining two main crystal axes. Thanks to the mirror symmetries relating the four CoBr$_4$ tetrahedra of a single unit cell, the field along $\mathbf{a}$ or $\mathbf{c}$ will be equivalent for all of them. If the anisotropy axis lies at angle $\beta$ with respect to the $\mathbf{a}$ direction  for a given Co$^{2+}$ ion, the resulting single-ion susceptibility would be $\chi^{\mathrm{SI}}_{\mathbf{a}}=\chi^{\mathrm{SI}}(\beta)$ and $\chi^{\mathrm{SI}}_{\mathbf{c}}=\chi^{\mathrm{SI}}(\pi/2-\beta)$. Susceptibility for a field, oriented at any angle, can easily be calculated numerically and the angle $\beta$ itself can be used as a fit parameter. Thus, the complete three-directional susceptibility data set is described by five parameters: anisotropy constant $D$, angle $\beta$ defining the orientation of the corresponding axis, and $g$ factors $g_{\alpha}$ with $\alpha=\mathbf{a},\mathbf{b},\mathbf{c}$. We can improve it further by taking into account the interactions between the ions on the mean-field level. This would require adding one extra mean-field parameter $J_{0}=\sum_{\pm\mathbf{r}}J_{\mathbf{r}}$, simply the sum of the relevant exchanges. The resulting mean-field susceptibility per ion for a given field direction would be

\begin{equation}\label{EQ:FullFitproIon}
  \chi^{\mathrm{MF}}_{\alpha}(T)=\dfrac{\chi^{\mathrm{SI}}_{\alpha}(T)}{1+\dfrac{J_{0}}{(g_{\alpha}\mu_{\mathrm{B}})^{2}}\chi^{\mathrm{SI}}_{\alpha}(T)}.
\end{equation}

And now, also taking care of the temperature-independent susceptibility contribution $\chi^{0}$, the final expression for the molar susceptibility reads

\begin{equation}\label{EQ:FullFit}
    \chi_{\alpha}(T)=N_{A}\chi^{\mathrm{MF}}_{\alpha}(T)+\chi^{0}_{\alpha}.
\end{equation}

Simultaneously fitting the data for all three field directions down to $20$~K (the mean-field approximation starts to break down below) we obtain the parameter estimates summarized in Table~\ref{TAB:Fit}.

\begin{table}
  \centering
  \begin{tabular}{c c l}
    \hline\hline
$D$ & & $14\pm1$~K\\
$\beta$ & & $44\pm1^{\circ}$\\
$J_{0}$ & & $5.5\pm0.2$~K\\
$g_{\mathbf{a}}$ & & $2.42\pm0.01$\\
$g_{\mathbf{b}}$ & & $2.47\pm0.02$\\
$g_{\mathbf{c}}$ & & $2.37\pm0.01$\\
$\chi^{0}_{\mathbf{a}}$ & & $5.8\pm1.7\cdot10^{-4}$~emu/mol\\
$\chi^{0}_{\mathbf{b}}$ & & $9.9\pm2.2\cdot10^{-4}$~emu/mol\\
$\chi^{0}_{\mathbf{c}}$ & & $4.2\pm1.7\cdot10^{-4}$~emu/mol\\
    \hline\hline
  \end{tabular}
    \caption{The results of the susceptibility fitting according to Eqs.~(\ref{EQ:FullFitproIon} , \ref{EQ:FullFit}).}\label{TAB:Fit}
\end{table}

\section{Torque data}
\label{APP:torque}

The two torque experiments reported in the present work are rather different in their sensitivity. This is mostly due to the fact that there is a dramatic difference in the sample mass, but also the details of the geometry might play a role. In the general scenario the torque experienced by the cantilever is given by the formula

\begin{equation}\label{EQ:torquegeneral}
  \mathbf{T}=[\mathbf{M}\times\mathbf{H}]+[\mathbf{L}\times(\mathbf{M}\cdot\nabla)\mathbf{H}].
\end{equation}

Here $\mathbf{L}$ is the distance from the fixed point of the cantilever to the sample position on it. Thus, both longitudinal and transverse (with respect to the field) components of magnetization may be contributing to the total cantilever deflection. The deflection is in turn measured as the change in capacitance of the device with the help of an Andeen-Hagerling 2550A bridge. However, depending on the shape the cantilever may be strongly resistant to bending or twisting in some ways, thus effectively excluding certain terms of Eq.~(\ref{EQ:torquegeneral}) from the game: the corresponding torque projection would not result in a measurable displacement.

In the $\mathbf{H}\parallel\mathbf{b}$ field experiment the large $15$~mg sample was used with the same setup as in Ref.~\cite{FengPovarov_PRB_2018_LinariteTilted}. The $\mathbf{b}$ axis of the sample was co-aligned with the simple one-leg cantilever direction. Taking into account the inevitable geometry imperfections, the resulting setup is effectively sensitive to all the magnetization components.

In contrast, the $\mathbf{H}\parallel\mathbf{a}$ measurement was performed in a Faraday balance setup~\cite{Blosser_RevSciInstr_2020_FaradayBalance}. This setup aims at optimizing the sensitivity to the longitudinal magnetization component, possibly eliminating contributions from the other ones. In addition, the geometry of the device urges one to use very small samples ($0.8$~mg in our case).

\section{Plateaux width toy model analysis}
\label{APP:toy}

\begin{figure}
  \centering
  \includegraphics[width=0.5\textwidth]{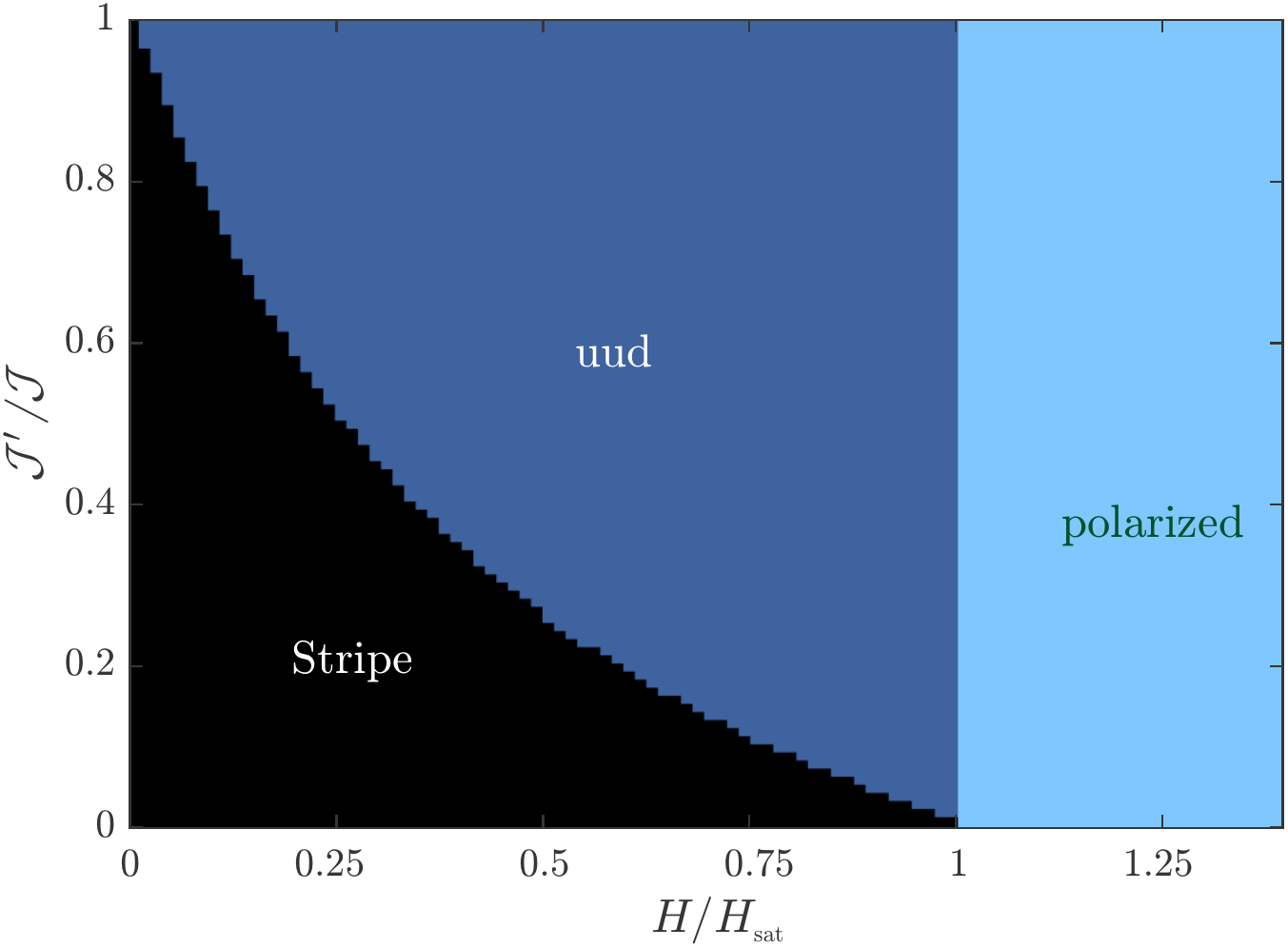}
  \caption{A ``phase diagram'' of the Eq.~(\ref{EQ:Isingtoy}) Ising toy model, consisting of three possible states: stripe, uud and polarized. The widths of $M=0$ and $M=1/3$ plateaux would be roughly equal to each other around $\mathcal{J}'/\mathcal{J}\sim0.3$.}\label{FIG:IsingDiag}
\end{figure}

In order to crudely estimate the possible $J'/J$ ratio, we would like to compare the width of the $M=0$ and $M=1/3$ plateaux. We do that by dealing with a toy Ising model that supports only the collinear states:

\begin{equation}\label{EQ:Isingtoy}
  \hamilt_{\text{Ising}}=\sum\limits_{i,j} \mathcal{J}\spin^{y}_{i,j}\spin^{y}_{i,j+1} + \mathcal{J}' \spin^{y}_{i,j}\spin^{y}_{i+1,j} + \mathcal{J}' \spin^{y}_{i,j}\spin^{y}_{i+1,j+1}
\end{equation}

The above Hamiltonian is the descendant of $\hamilt_{1/2}$ (Eq.~\ref{EQ:XYZHamilt}) from the main text with $4J\rightarrow \mathcal{J}$, $4J'\rightarrow \mathcal{J}'$, and $x,z$ spin components truncated. \textcolor[rgb]{0.00,0.00,0.00}{It can be seen as somewhat relevant to the material $\alpha$-CoV$_2$O$_6$~\cite{HeYamaura_JACS_2009_CoV2O6plateus} discussed in the main text.}

 We fix the sum $\mathcal{J}+2\mathcal{J}'$ such that it would give the constant saturation field $H_{\mathrm{sat}}$, and then we investigate which of the two possible structures provides the minimal energy in a given magnetic field. The resulting ``phase diagram'' as a function of the magnetic field and the exchange ratio is given in Fig.~\ref{FIG:IsingDiag}. One can see that the uud phase is absent for vanishing $\mathcal{J}'/\mathcal{J}$, and it extends down to zero field in the nondistorted triangular lattice limit, in agreement with the well-known results~\cite{Yamamoto_PRL_2014_TriangLatXXZ,ChenJuJiang_PRB_2013_DistortTriangquasi1D}.

In the case of \CCoB\ the widths of the stripe and uud phases are roughly equal to each other. Thus, from this very naive toy model we can make a ``zeroth-order'' estimate for the frustration ratio $J'/J\sim0.3$. This renders \CCoB\ as a quasi-2D rather than quasi-1D dimensional material from the coupling strength point of view. In the notations of the original $\hamilt_{3/2}$ Hamiltonian those couplings are $J\sim1.5$ and $J'\sim0.5$~K.

\bibliography{d:/The_Library}

\begin{thebibliography}{57}%
\makeatletter
\providecommand \@ifxundefined [1]{%
 \@ifx{#1\undefined}
}%
\providecommand \@ifnum [1]{%
 \ifnum #1\expandafter \@firstoftwo
 \else \expandafter \@secondoftwo
 \fi
}%
\providecommand \@ifx [1]{%
 \ifx #1\expandafter \@firstoftwo
 \else \expandafter \@secondoftwo
 \fi
}%
\providecommand \natexlab [1]{#1}%
\providecommand \enquote  [1]{``#1''}%
\providecommand \bibnamefont  [1]{#1}%
\providecommand \bibfnamefont [1]{#1}%
\providecommand \citenamefont [1]{#1}%
\providecommand \href@noop [0]{\@secondoftwo}%
\providecommand \href [0]{\begingroup \@sanitize@url \@href}%
\providecommand \@href[1]{\@@startlink{#1}\@@href}%
\providecommand \@@href[1]{\endgroup#1\@@endlink}%
\providecommand \@sanitize@url [0]{\catcode `\\12\catcode `\$12\catcode
  `\&12\catcode `\#12\catcode `\^12\catcode `\_12\catcode `\%12\relax}%
\providecommand \@@startlink[1]{}%
\providecommand \@@endlink[0]{}%
\providecommand \url  [0]{\begingroup\@sanitize@url \@url }%
\providecommand \@url [1]{\endgroup\@href {#1}{\urlprefix }}%
\providecommand \urlprefix  [0]{URL }%
\providecommand \Eprint [0]{\href }%
\providecommand \doibase [0]{http://dx.doi.org/}%
\providecommand \selectlanguage [0]{\@gobble}%
\providecommand \bibinfo  [0]{\@secondoftwo}%
\providecommand \bibfield  [0]{\@secondoftwo}%
\providecommand \translation [1]{[#1]}%
\providecommand \BibitemOpen [0]{}%
\providecommand \bibitemStop [0]{}%
\providecommand \bibitemNoStop [0]{.\EOS\space}%
\providecommand \EOS [0]{\spacefactor3000\relax}%
\providecommand \BibitemShut  [1]{\csname bibitem#1\endcsname}%
\let\auto@bib@innerbib\@empty
%</preamble>
\bibitem [{\citenamefont
  {Starykh}(2015)}]{Starykh_RepPrPhys_2015_TriangularReview}%
  \BibitemOpen
  \bibfield  {author} {\bibinfo {author} {\bibfnamefont {O.~A.}\ \bibnamefont
  {Starykh}},\ }\bibfield  {title} {\enquote {\bibinfo {title} {{Unusual
  ordered phases of highly frustrated magnets: a review}},}\ }\href {\doibase
  10.1088/0034-4885/78/5/052502} {\bibfield  {journal} {\bibinfo  {journal}
  {Rep. Prog. Phys.}\ }\textbf {\bibinfo {volume} {78}},\ \bibinfo {pages}
  {052502} (\bibinfo {year} {2015})}\BibitemShut {NoStop}%
\bibitem [{\citenamefont {Yamamoto}\ \emph {et~al.}(2014)\citenamefont
  {Yamamoto}, \citenamefont {Marmorini},\ and\ \citenamefont
  {Danshita}}]{Yamamoto_PRL_2014_TriangLatXXZ}%
  \BibitemOpen
  \bibfield  {author} {\bibinfo {author} {\bibfnamefont {D.}~\bibnamefont
  {Yamamoto}}, \bibinfo {author} {\bibfnamefont {G.}~\bibnamefont {Marmorini}},
  \ and\ \bibinfo {author} {\bibfnamefont {I.}~\bibnamefont {Danshita}},\
  }\bibfield  {title} {\enquote {\bibinfo {title} {{Quantum Phase Diagram of
  the Triangular-Lattice $XXZ$ Model in a Magnetic Field}},}\ }\href {\doibase
  10.1103/PhysRevLett.112.127203} {\bibfield  {journal} {\bibinfo  {journal}
  {Phys. Rev. Lett.}\ }\textbf {\bibinfo {volume} {112}},\ \bibinfo {pages}
  {127203} (\bibinfo {year} {2014})}\BibitemShut {NoStop}%
\bibitem [{\citenamefont {Sellmann}\ \emph {et~al.}(2015)\citenamefont
  {Sellmann}, \citenamefont {Zhang},\ and\ \citenamefont
  {Eggert}}]{SellmannZhangEggert_PRB_2015_AnotherXXZtriangularPhD}%
  \BibitemOpen
  \bibfield  {author} {\bibinfo {author} {\bibfnamefont {D.}~\bibnamefont
  {Sellmann}}, \bibinfo {author} {\bibfnamefont {X.-F.}\ \bibnamefont {Zhang}},
  \ and\ \bibinfo {author} {\bibfnamefont {S.}~\bibnamefont {Eggert}},\
  }\bibfield  {title} {\enquote {\bibinfo {title} {{Phase diagram of the
  antiferromagnetic XXZ model on the triangular lattice}},}\ }\href {\doibase
  10.1103/PhysRevB.91.081104} {\bibfield  {journal} {\bibinfo  {journal} {Phys.
  Rev. B}\ }\textbf {\bibinfo {volume} {91}},\ \bibinfo {pages} {081104}
  (\bibinfo {year} {2015})}\BibitemShut {NoStop}%
\bibitem [{\citenamefont {Ross}\ \emph {et~al.}(2011)\citenamefont {Ross},
  \citenamefont {Savary}, \citenamefont {Gaulin},\ and\ \citenamefont
  {Balents}}]{RossSavary_PRX_2011_QuantumIceNeutrons}%
  \BibitemOpen
  \bibfield  {author} {\bibinfo {author} {\bibfnamefont {K.~A.}\ \bibnamefont
  {Ross}}, \bibinfo {author} {\bibfnamefont {L.}~\bibnamefont {Savary}},
  \bibinfo {author} {\bibfnamefont {B.~D.}\ \bibnamefont {Gaulin}}, \ and\
  \bibinfo {author} {\bibfnamefont {L.}~\bibnamefont {Balents}},\ }\bibfield
  {title} {\enquote {\bibinfo {title} {{Quantum Excitations in Quantum Spin
  Ice}},}\ }\href {\doibase 10.1103/PhysRevX.1.021002} {\bibfield  {journal}
  {\bibinfo  {journal} {Phys. Rev. X}\ }\textbf {\bibinfo {volume} {1}},\
  \bibinfo {pages} {021002} (\bibinfo {year} {2011})}\BibitemShut {NoStop}%
\bibitem [{\citenamefont {Taillefumier}\ \emph {et~al.}(2017)\citenamefont
  {Taillefumier}, \citenamefont {Benton}, \citenamefont {Yan}, \citenamefont
  {Jaubert},\ and\ \citenamefont
  {Shannon}}]{TaillefumierBenton_PRX_2017_QSicePhases}%
  \BibitemOpen
  \bibfield  {author} {\bibinfo {author} {\bibfnamefont {M.}~\bibnamefont
  {Taillefumier}}, \bibinfo {author} {\bibfnamefont {O.}~\bibnamefont
  {Benton}}, \bibinfo {author} {\bibfnamefont {H.}~\bibnamefont {Yan}},
  \bibinfo {author} {\bibfnamefont {L.~D.~C.}\ \bibnamefont {Jaubert}}, \ and\
  \bibinfo {author} {\bibfnamefont {N.}~\bibnamefont {Shannon}},\ }\bibfield
  {title} {\enquote {\bibinfo {title} {{Competing Spin Liquids and Hidden
  Spin-Nematic Order in Spin Ice with Frustrated Transverse Exchange}},}\
  }\href {\doibase 10.1103/PhysRevX.7.041057} {\bibfield  {journal} {\bibinfo
  {journal} {Phys. Rev. X}\ }\textbf {\bibinfo {volume} {7}},\ \bibinfo {pages}
  {041057} (\bibinfo {year} {2017})}\BibitemShut {NoStop}%
\bibitem [{\citenamefont {Gingras}\ and\ \citenamefont
  {McClarty}(2014)}]{GingrasMcClarty_RepProgPhys_2014_QSIreview}%
  \BibitemOpen
  \bibfield  {author} {\bibinfo {author} {\bibfnamefont {M.~J.~P.}\
  \bibnamefont {Gingras}}\ and\ \bibinfo {author} {\bibfnamefont {P.~A.}\
  \bibnamefont {McClarty}},\ }\bibfield  {title} {\enquote {\bibinfo {title}
  {{Quantum spin ice: a search for gapless quantum spin liquids in pyrochlore
  magnets}},}\ }\href {\doibase 10.1088/0034-4885/77/5/056501} {\bibfield
  {journal} {\bibinfo  {journal} {Rep. Prog. Phys.}\ }\textbf {\bibinfo
  {volume} {77}},\ \bibinfo {pages} {056501} (\bibinfo {year}
  {2014})}\BibitemShut {NoStop}%
\bibitem [{\citenamefont {Kitaev}(2006)}]{Kitaev_AnnPhys_2006_Kitaev}%
  \BibitemOpen
  \bibfield  {author} {\bibinfo {author} {\bibfnamefont {A.}~\bibnamefont
  {Kitaev}},\ }\bibfield  {title} {\enquote {\bibinfo {title} {Anyons in an
  exactly solved model and beyond},}\ }\href {\doibase
  https://doi.org/10.1016/j.aop.2005.10.005} {\bibfield  {journal} {\bibinfo
  {journal} {Ann. Phys.}\ }\textbf {\bibinfo {volume} {321}},\ \bibinfo {pages}
  {2} (\bibinfo {year} {2006})}\BibitemShut {NoStop}%
\bibitem [{\citenamefont {M.}\ \emph {et~al.}(2017)\citenamefont {M.},
  \citenamefont {Tsirlin}, \citenamefont {Daghofer}, \citenamefont {van~den
  Brink}, \citenamefont {Singh}, \citenamefont {Gegenwart},\ and\ \citenamefont
  {Valent{\'{\i}}}}]{WinterTsirlin_JPCM_2017_KitaevReview}%
  \BibitemOpen
  \bibfield  {author} {\bibinfo {author} {\bibfnamefont {Winter~S.}\
  \bibnamefont {M.}}, \bibinfo {author} {\bibfnamefont {A.~A.}\ \bibnamefont
  {Tsirlin}}, \bibinfo {author} {\bibfnamefont {M.}~\bibnamefont {Daghofer}},
  \bibinfo {author} {\bibfnamefont {J.}~\bibnamefont {van~den Brink}}, \bibinfo
  {author} {\bibfnamefont {Y.}~\bibnamefont {Singh}}, \bibinfo {author}
  {\bibfnamefont {P.}~\bibnamefont {Gegenwart}}, \ and\ \bibinfo {author}
  {\bibfnamefont {R.}~\bibnamefont {Valent{\'{\i}}}},\ }\bibfield  {title}
  {\enquote {\bibinfo {title} {{Models and materials for generalized Kitaev
  magnetism}},}\ }\href {\doibase 10.1088/1361-648x/aa8cf5} {\bibfield
  {journal} {\bibinfo  {journal} {J. Phys.: Cond. Mat.}\ }\textbf {\bibinfo
  {volume} {29}},\ \bibinfo {pages} {493002} (\bibinfo {year}
  {2017})}\BibitemShut {NoStop}%
\bibitem [{\citenamefont {Takagi}\ \emph {et~al.}(2019)\citenamefont {Takagi},
  \citenamefont {Takayama}, \citenamefont {Jackeli}, \citenamefont
  {Khaliullin},\ and\ \citenamefont
  {Nagler}}]{TakagiTakayama_NatRevPhys_2019_KitaevReview}%
  \BibitemOpen
  \bibfield  {author} {\bibinfo {author} {\bibfnamefont {H.}~\bibnamefont
  {Takagi}}, \bibinfo {author} {\bibfnamefont {T.}~\bibnamefont {Takayama}},
  \bibinfo {author} {\bibfnamefont {G.}~\bibnamefont {Jackeli}}, \bibinfo
  {author} {\bibfnamefont {G.}~\bibnamefont {Khaliullin}}, \ and\ \bibinfo
  {author} {\bibfnamefont {S.~E.}\ \bibnamefont {Nagler}},\ }\bibfield  {title}
  {\enquote {\bibinfo {title} {{Concept and realization of Kitaev quantum spin
  liquids}},}\ }\href {\doibase 10.1038/s42254-019-0038-2} {\bibfield
  {journal} {\bibinfo  {journal} {Nat. Rev. Phys.}\ }\textbf {\bibinfo {volume}
  {1}},\ \bibinfo {pages} {264} (\bibinfo {year} {2019})}\BibitemShut {NoStop}%
\bibitem [{\citenamefont {Liu}\ \emph {et~al.}(2020)\citenamefont {Liu},
  \citenamefont {Chaloupka},\ and\ \citenamefont
  {Khaliullin}}]{LiuChaloupkaKhaliullin_PRL_2020_CobaltKitaev}%
  \BibitemOpen
  \bibfield  {author} {\bibinfo {author} {\bibfnamefont {H.}~\bibnamefont
  {Liu}}, \bibinfo {author} {\bibfnamefont {J.}~\bibnamefont {Chaloupka}}, \
  and\ \bibinfo {author} {\bibfnamefont {G.}~\bibnamefont {Khaliullin}},\
  }\bibfield  {title} {\enquote {\bibinfo {title} {Kitaev spin liquid in $3d$
  transition metal compounds},}\ }\href {\doibase
  10.1103/PhysRevLett.125.047201} {\bibfield  {journal} {\bibinfo  {journal}
  {Phys. Rev. Lett.}\ }\textbf {\bibinfo {volume} {125}},\ \bibinfo {pages}
  {047201} (\bibinfo {year} {2020})}\BibitemShut {NoStop}%
\bibitem [{\citenamefont {Zhu}\ \emph {et~al.}(2018)\citenamefont {Zhu},
  \citenamefont {Maksimov}, \citenamefont {White},\ and\ \citenamefont
  {Chernyshev}}]{ZhuMaksimov_PRL_2018_AnisotropicTriangLattice}%
  \BibitemOpen
  \bibfield  {author} {\bibinfo {author} {\bibfnamefont {Z.}~\bibnamefont
  {Zhu}}, \bibinfo {author} {\bibfnamefont {P.~A.}\ \bibnamefont {Maksimov}},
  \bibinfo {author} {\bibfnamefont {S.~R.}\ \bibnamefont {White}}, \ and\
  \bibinfo {author} {\bibfnamefont {A.~L.}\ \bibnamefont {Chernyshev}},\
  }\bibfield  {title} {\enquote {\bibinfo {title} {{Topography of Spin Liquids
  on a Triangular Lattice}},}\ }\href {\doibase 10.1103/PhysRevLett.120.207203}
  {\bibfield  {journal} {\bibinfo  {journal} {Phys. Rev. Lett.}\ }\textbf
  {\bibinfo {volume} {120}},\ \bibinfo {pages} {207203} (\bibinfo {year}
  {2018})}\BibitemShut {NoStop}%
\bibitem [{\citenamefont {Maksimov}\ \emph {et~al.}(2019)\citenamefont
  {Maksimov}, \citenamefont {Zhu}, \citenamefont {White},\ and\ \citenamefont
  {Chernyshev}}]{MaksimovZhu_PRX_2019_AnisotropicTriangLattice}%
  \BibitemOpen
  \bibfield  {author} {\bibinfo {author} {\bibfnamefont {P.~A.}\ \bibnamefont
  {Maksimov}}, \bibinfo {author} {\bibfnamefont {Z.}~\bibnamefont {Zhu}},
  \bibinfo {author} {\bibfnamefont {S.~R.}\ \bibnamefont {White}}, \ and\
  \bibinfo {author} {\bibfnamefont {A.~L.}\ \bibnamefont {Chernyshev}},\
  }\bibfield  {title} {\enquote {\bibinfo {title} {{Anisotropic-Exchange
  Magnets on a Triangular Lattice: Spin Waves, Accidental Degeneracies, and
  Dual Spin Liquids}},}\ }\href {\doibase 10.1103/PhysRevX.9.021017} {\bibfield
   {journal} {\bibinfo  {journal} {Phys. Rev. X}\ }\textbf {\bibinfo {volume}
  {9}},\ \bibinfo {pages} {021017} (\bibinfo {year} {2019})}\BibitemShut
  {NoStop}%
\bibitem [{\citenamefont {Coldea}\ \emph {et~al.}(2001)\citenamefont {Coldea},
  \citenamefont {Tennant}, \citenamefont {Tsvelik},\ and\ \citenamefont
  {Tylczynski}}]{Coldea_PRL_2001_CCC2D}%
  \BibitemOpen
  \bibfield  {author} {\bibinfo {author} {\bibfnamefont {R.}~\bibnamefont
  {Coldea}}, \bibinfo {author} {\bibfnamefont {D.~A.}\ \bibnamefont {Tennant}},
  \bibinfo {author} {\bibfnamefont {A.~M.}\ \bibnamefont {Tsvelik}}, \ and\
  \bibinfo {author} {\bibfnamefont {Z.}~\bibnamefont {Tylczynski}},\ }\bibfield
   {title} {\enquote {\bibinfo {title} {{Experimental Realization of a 2D
  Fractional Quantum Spin Liquid}},}\ }\href {\doibase
  10.1103/PhysRevLett.86.1335} {\bibfield  {journal} {\bibinfo  {journal}
  {Phys. Rev. Lett.}\ }\textbf {\bibinfo {volume} {86}},\ \bibinfo {pages}
  {1335} (\bibinfo {year} {2001})}\BibitemShut {NoStop}%
\bibitem [{\citenamefont {Tokiwa}\ \emph {et~al.}(2006)\citenamefont {Tokiwa},
  \citenamefont {Radu}, \citenamefont {Coldea}, \citenamefont {Wilhelm},
  \citenamefont {Tylczynski},\ and\ \citenamefont
  {Steglich}}]{Tokiwa_PRB_2006_Cs2CuCl4phases}%
  \BibitemOpen
  \bibfield  {author} {\bibinfo {author} {\bibfnamefont {Y.}~\bibnamefont
  {Tokiwa}}, \bibinfo {author} {\bibfnamefont {T.}~\bibnamefont {Radu}},
  \bibinfo {author} {\bibfnamefont {R.}~\bibnamefont {Coldea}}, \bibinfo
  {author} {\bibfnamefont {H.}~\bibnamefont {Wilhelm}}, \bibinfo {author}
  {\bibfnamefont {Z.}~\bibnamefont {Tylczynski}}, \ and\ \bibinfo {author}
  {\bibfnamefont {F.}~\bibnamefont {Steglich}},\ }\bibfield  {title} {\enquote
  {\bibinfo {title} {{Magnetic phase transitions in the two-dimensional
  frustrated quantum antiferromagnet
  ${\mathrm{Cs}}_{2}\mathrm{Cu}{\mathrm{Cl}}_{4}$}},}\ }\href {\doibase
  10.1103/PhysRevB.73.134414} {\bibfield  {journal} {\bibinfo  {journal} {Phys.
  Rev. B}\ }\textbf {\bibinfo {volume} {73}},\ \bibinfo {pages} {134414}
  (\bibinfo {year} {2006})}\BibitemShut {NoStop}%
\bibitem [{\citenamefont {Smirnov}\ \emph {et~al.}(2012)\citenamefont
  {Smirnov}, \citenamefont {Povarov}, \citenamefont {Petrov},\ and\
  \citenamefont {Shapiro}}]{SmirnovPovarov_PRB_2012_ESRordered}%
  \BibitemOpen
  \bibfield  {author} {\bibinfo {author} {\bibfnamefont {A.~I.}\ \bibnamefont
  {Smirnov}}, \bibinfo {author} {\bibfnamefont {{\relax K. Yu.}}~\bibnamefont
  {Povarov}}, \bibinfo {author} {\bibfnamefont {S.~V.}\ \bibnamefont {Petrov}},
  \ and\ \bibinfo {author} {\bibfnamefont {{\relax A. Ya.}}~\bibnamefont
  {Shapiro}},\ }\bibfield  {title} {\enquote {\bibinfo {title} {{Magnetic
  resonance in the ordered phases of the two-dimensional frustrated quantum
  magnet Cs${}_{2}$CuCl${}_{4}$}},}\ }\href {\doibase
  10.1103/PhysRevB.85.184423} {\bibfield  {journal} {\bibinfo  {journal} {Phys.
  Rev. B}\ }\textbf {\bibinfo {volume} {85}},\ \bibinfo {pages} {184423}
  (\bibinfo {year} {2012})}\BibitemShut {NoStop}%
\bibitem [{\citenamefont {Schulze}\ \emph {et~al.}(2019)\citenamefont
  {Schulze}, \citenamefont {Arsenijevic}, \citenamefont {Opherden},
  \citenamefont {Ponomaryov}, \citenamefont {Wosnitza}, \citenamefont {Ono},
  \citenamefont {Tanaka},\ and\ \citenamefont
  {Zvyagin}}]{SchulzeArsenijevic_PRR_2019_CCuC1Dheattrans}%
  \BibitemOpen
  \bibfield  {author} {\bibinfo {author} {\bibfnamefont {E.}~\bibnamefont
  {Schulze}}, \bibinfo {author} {\bibfnamefont {S.}~\bibnamefont
  {Arsenijevic}}, \bibinfo {author} {\bibfnamefont {L.}~\bibnamefont
  {Opherden}}, \bibinfo {author} {\bibfnamefont {A.~N.}\ \bibnamefont
  {Ponomaryov}}, \bibinfo {author} {\bibfnamefont {J.}~\bibnamefont
  {Wosnitza}}, \bibinfo {author} {\bibfnamefont {T.}~\bibnamefont {Ono}},
  \bibinfo {author} {\bibfnamefont {H.}~\bibnamefont {Tanaka}}, \ and\ \bibinfo
  {author} {\bibfnamefont {S.~A.}\ \bibnamefont {Zvyagin}},\ }\bibfield
  {title} {\enquote {\bibinfo {title} {{Evidence of one-dimensional magnetic
  heat transport in the triangular-lattice antiferromagnet
  ${\mathrm{Cs}}_{2}{\mathrm{CuCl}}_{4}$}},}\ }\href {\doibase
  10.1103/PhysRevResearch.1.032022} {\bibfield  {journal} {\bibinfo  {journal}
  {Phys. Rev. Research}\ }\textbf {\bibinfo {volume} {1}},\ \bibinfo {pages}
  {032022} (\bibinfo {year} {2019})}\BibitemShut {NoStop}%
\bibitem [{\citenamefont {Starykh}\ \emph {et~al.}(2010)\citenamefont
  {Starykh}, \citenamefont {Katsura},\ and\ \citenamefont
  {Balents}}]{Starykh_PRB_2010_Cs2CuCl4theory}%
  \BibitemOpen
  \bibfield  {author} {\bibinfo {author} {\bibfnamefont {O.~A.}\ \bibnamefont
  {Starykh}}, \bibinfo {author} {\bibfnamefont {H.}~\bibnamefont {Katsura}}, \
  and\ \bibinfo {author} {\bibfnamefont {L.}~\bibnamefont {Balents}},\
  }\bibfield  {title} {\enquote {\bibinfo {title} {{Extreme sensitivity of a
  frustrated quantum magnet:
  ${\mathrm{Cs}}_{2}\mathrm{Cu}{\mathrm{Cl}}_{4}$}},}\ }\href {\doibase
  10.1103/PhysRevB.82.014421} {\bibfield  {journal} {\bibinfo  {journal} {Phys.
  Rev. B}\ }\textbf {\bibinfo {volume} {82}},\ \bibinfo {pages} {014421}
  (\bibinfo {year} {2010})}\BibitemShut {NoStop}%
\bibitem [{\citenamefont {Figgis}\ \emph {et~al.}(1987)\citenamefont {Figgis},
  \citenamefont {Reynolds},\ and\ \citenamefont
  {White}}]{Figgis_JChemSoc_1987_Cs2CoCl4refinement}%
  \BibitemOpen
  \bibfield  {author} {\bibinfo {author} {\bibfnamefont {B.~N.}\ \bibnamefont
  {Figgis}}, \bibinfo {author} {\bibfnamefont {P.~A.}\ \bibnamefont
  {Reynolds}}, \ and\ \bibinfo {author} {\bibfnamefont {A.~H.}\ \bibnamefont
  {White}},\ }\bibfield  {title} {\enquote {\bibinfo {title} {{Charge density
  in the CoCl$_{4}^{2-}$ ion: a comparison with spin density and theoretical
  calculations}},}\ }\href {\doibase 10.1039/dt9870001737} {\bibfield
  {journal} {\bibinfo  {journal} {J. Chem. Soc., Dalton Trans.}\ ,\ \bibinfo
  {pages} {1737}} (\bibinfo {year} {1987})}\BibitemShut {NoStop}%
\bibitem [{\citenamefont {Kenzelmann}\ \emph {et~al.}(2002)\citenamefont
  {Kenzelmann}, \citenamefont {Coldea}, \citenamefont {Tennant}, \citenamefont
  {Visser}, \citenamefont {Hofmann}, \citenamefont {Smeibidl},\ and\
  \citenamefont {Tylczynski}}]{KenzelmannColdea_PRB_2002_CsCoCldiffraction}%
  \BibitemOpen
  \bibfield  {author} {\bibinfo {author} {\bibfnamefont {M.}~\bibnamefont
  {Kenzelmann}}, \bibinfo {author} {\bibfnamefont {R.}~\bibnamefont {Coldea}},
  \bibinfo {author} {\bibfnamefont {D.~A.}\ \bibnamefont {Tennant}}, \bibinfo
  {author} {\bibfnamefont {D.}~\bibnamefont {Visser}}, \bibinfo {author}
  {\bibfnamefont {M.}~\bibnamefont {Hofmann}}, \bibinfo {author} {\bibfnamefont
  {P.}~\bibnamefont {Smeibidl}}, \ and\ \bibinfo {author} {\bibfnamefont
  {Z.}~\bibnamefont {Tylczynski}},\ }\bibfield  {title} {\enquote {\bibinfo
  {title} {{Order-to-disorder transition in the $\mathrm{XY}$-like quantum
  magnet ${\mathrm{Cs}}_{2}{\mathrm{CoCl}}_{4}$ induced by noncommuting applied
  fields}},}\ }\href {\doibase 10.1103/PhysRevB.65.144432} {\bibfield
  {journal} {\bibinfo  {journal} {Phys. Rev. B}\ }\textbf {\bibinfo {volume}
  {65}},\ \bibinfo {pages} {144432} (\bibinfo {year} {2002})}\BibitemShut
  {NoStop}%
\bibitem [{\citenamefont {Breunig}\ \emph {et~al.}(2013)\citenamefont
  {Breunig}, \citenamefont {Garst}, \citenamefont {Sela}, \citenamefont
  {Buldmann}, \citenamefont {Becker}, \citenamefont {Bohat\'y}, \citenamefont
  {M\"uller},\ and\ \citenamefont {Lorenz}}]{BreunigGarst_PRL_2013_CsCoCl_TF}%
  \BibitemOpen
  \bibfield  {author} {\bibinfo {author} {\bibfnamefont {O.}~\bibnamefont
  {Breunig}}, \bibinfo {author} {\bibfnamefont {M.}~\bibnamefont {Garst}},
  \bibinfo {author} {\bibfnamefont {E.}~\bibnamefont {Sela}}, \bibinfo {author}
  {\bibfnamefont {B.}~\bibnamefont {Buldmann}}, \bibinfo {author}
  {\bibfnamefont {P.}~\bibnamefont {Becker}}, \bibinfo {author} {\bibfnamefont
  {L.}~\bibnamefont {Bohat\'y}}, \bibinfo {author} {\bibfnamefont
  {R.}~\bibnamefont {M\"uller}}, \ and\ \bibinfo {author} {\bibfnamefont
  {T.}~\bibnamefont {Lorenz}},\ }\bibfield  {title} {\enquote {\bibinfo {title}
  {{Spin-$\frac{1}{2}$ $XXZ$ Chain System
  ${\mathrm{Cs}}_{2}{\mathrm{CoCl}}_{4}$ in a Transverse Magnetic Field}},}\
  }\href {\doibase 10.1103/PhysRevLett.111.187202} {\bibfield  {journal}
  {\bibinfo  {journal} {Phys. Rev. Lett.}\ }\textbf {\bibinfo {volume} {111}},\
  \bibinfo {pages} {187202} (\bibinfo {year} {2013})}\BibitemShut {NoStop}%
\bibitem [{\citenamefont {Breunig}\ \emph {et~al.}(2015)\citenamefont
  {Breunig}, \citenamefont {Garst}, \citenamefont {Rosch}, \citenamefont
  {Sela}, \citenamefont {Buldmann}, \citenamefont {Becker}, \citenamefont
  {Bohat\'y}, \citenamefont {M\"uller},\ and\ \citenamefont
  {Lorenz}}]{Breunig_PRB_2015_CsCoClPhD}%
  \BibitemOpen
  \bibfield  {author} {\bibinfo {author} {\bibfnamefont {O.}~\bibnamefont
  {Breunig}}, \bibinfo {author} {\bibfnamefont {M.}~\bibnamefont {Garst}},
  \bibinfo {author} {\bibfnamefont {A.}~\bibnamefont {Rosch}}, \bibinfo
  {author} {\bibfnamefont {E.}~\bibnamefont {Sela}}, \bibinfo {author}
  {\bibfnamefont {B.}~\bibnamefont {Buldmann}}, \bibinfo {author}
  {\bibfnamefont {P.}~\bibnamefont {Becker}}, \bibinfo {author} {\bibfnamefont
  {L.}~\bibnamefont {Bohat\'y}}, \bibinfo {author} {\bibfnamefont
  {R.}~\bibnamefont {M\"uller}}, \ and\ \bibinfo {author} {\bibfnamefont
  {T.}~\bibnamefont {Lorenz}},\ }\bibfield  {title} {\enquote {\bibinfo {title}
  {{Low-temperature ordered phases of the spin-$\frac{1}{2}$ XXZ chain system
  ${\mathrm{Cs}}_{2}{\mathrm{CoCl}}_{4}$}},}\ }\href {\doibase
  10.1103/PhysRevB.91.024423} {\bibfield  {journal} {\bibinfo  {journal} {Phys.
  Rev. B}\ }\textbf {\bibinfo {volume} {91}},\ \bibinfo {pages} {024423}
  (\bibinfo {year} {2015})}\BibitemShut {NoStop}%
\bibitem [{\citenamefont {Ono}\ \emph {et~al.}(2003)\citenamefont {Ono},
  \citenamefont {Tanaka}, \citenamefont {Aruga~Katori}, \citenamefont
  {Ishikawa}, \citenamefont {Mitamura},\ and\ \citenamefont
  {Goto}}]{OnoTanaka_PRB_2003_CCBplateau}%
  \BibitemOpen
  \bibfield  {author} {\bibinfo {author} {\bibfnamefont {T.}~\bibnamefont
  {Ono}}, \bibinfo {author} {\bibfnamefont {H.}~\bibnamefont {Tanaka}},
  \bibinfo {author} {\bibfnamefont {H.}~\bibnamefont {Aruga~Katori}}, \bibinfo
  {author} {\bibfnamefont {F.}~\bibnamefont {Ishikawa}}, \bibinfo {author}
  {\bibfnamefont {H.}~\bibnamefont {Mitamura}}, \ and\ \bibinfo {author}
  {\bibfnamefont {T.}~\bibnamefont {Goto}},\ }\bibfield  {title} {\enquote
  {\bibinfo {title} {{Magnetization plateau in the frustrated quantum spin
  system ${\mathrm{Cs}}_{2}{\mathrm{CuBr}}_{4}$}},}\ }\href {\doibase
  10.1103/PhysRevB.67.104431} {\bibfield  {journal} {\bibinfo  {journal} {Phys.
  Rev. B}\ }\textbf {\bibinfo {volume} {67}},\ \bibinfo {pages} {104431}
  (\bibinfo {year} {2003})}\BibitemShut {NoStop}%
\bibitem [{\citenamefont {Tsujii}\ \emph {et~al.}(2007)\citenamefont {Tsujii},
  \citenamefont {Rotundu}, \citenamefont {Ono}, \citenamefont {Tanaka},
  \citenamefont {Andraka}, \citenamefont {Ingersent},\ and\ \citenamefont
  {Takano}}]{TsujiiRotundu_PRB_2007_Cs2CuBr4UUD}%
  \BibitemOpen
  \bibfield  {author} {\bibinfo {author} {\bibfnamefont {H.}~\bibnamefont
  {Tsujii}}, \bibinfo {author} {\bibfnamefont {C.~R.}\ \bibnamefont {Rotundu}},
  \bibinfo {author} {\bibfnamefont {T.}~\bibnamefont {Ono}}, \bibinfo {author}
  {\bibfnamefont {H.}~\bibnamefont {Tanaka}}, \bibinfo {author} {\bibfnamefont
  {B.}~\bibnamefont {Andraka}}, \bibinfo {author} {\bibfnamefont
  {K.}~\bibnamefont {Ingersent}}, \ and\ \bibinfo {author} {\bibfnamefont
  {Y.}~\bibnamefont {Takano}},\ }\bibfield  {title} {\enquote {\bibinfo {title}
  {{Thermodynamics of the up-up-down phase of the $S=\frac{1}{2}$
  triangular-lattice antiferromagnet
  ${\mathrm{Cs}}_{2}\mathrm{Cu}{\mathrm{Br}}_{4}$}},}\ }\href {\doibase
  10.1103/PhysRevB.76.060406} {\bibfield  {journal} {\bibinfo  {journal} {Phys.
  Rev. B}\ }\textbf {\bibinfo {volume} {76}},\ \bibinfo {pages} {060406}
  (\bibinfo {year} {2007})}\BibitemShut {NoStop}%
\bibitem [{\citenamefont {Fortune}\ \emph {et~al.}(2009)\citenamefont
  {Fortune}, \citenamefont {Hannahs}, \citenamefont {Yoshida}, \citenamefont
  {Sherline}, \citenamefont {Ono}, \citenamefont {Tanaka},\ and\ \citenamefont
  {Takano}}]{FortuneHannahs_PRL_2009_CCBtransitions}%
  \BibitemOpen
  \bibfield  {author} {\bibinfo {author} {\bibfnamefont {N.~A.}\ \bibnamefont
  {Fortune}}, \bibinfo {author} {\bibfnamefont {S.~T.}\ \bibnamefont
  {Hannahs}}, \bibinfo {author} {\bibfnamefont {Y.}~\bibnamefont {Yoshida}},
  \bibinfo {author} {\bibfnamefont {T.~E.}\ \bibnamefont {Sherline}}, \bibinfo
  {author} {\bibfnamefont {T.}~\bibnamefont {Ono}}, \bibinfo {author}
  {\bibfnamefont {H.}~\bibnamefont {Tanaka}}, \ and\ \bibinfo {author}
  {\bibfnamefont {Y.}~\bibnamefont {Takano}},\ }\bibfield  {title} {\enquote
  {\bibinfo {title} {{Cascade of Magnetic-Field-Induced Quantum Phase
  Transitions in a Spin-$\frac{1}{2}$ Triangular-Lattice Antiferromagnet}},}\
  }\href {\doibase 10.1103/PhysRevLett.102.257201} {\bibfield  {journal}
  {\bibinfo  {journal} {Phys. Rev. Lett.}\ }\textbf {\bibinfo {volume} {102}},\
  \bibinfo {pages} {257201} (\bibinfo {year} {2009})}\BibitemShut {NoStop}%
\bibitem [{\citenamefont {Seifert}\ and\ \citenamefont
  {Al-Khudair}(1975)}]{SeifertKhudair_JInorgNucChem_1975_A2CoX4bridgeman}%
  \BibitemOpen
  \bibfield  {author} {\bibinfo {author} {\bibfnamefont {H.~J.}\ \bibnamefont
  {Seifert}}\ and\ \bibinfo {author} {\bibfnamefont {I.}~\bibnamefont
  {Al-Khudair}},\ }\bibfield  {title} {\enquote {\bibinfo {title} {{\"{U}ber
  die systeme alkalimetallbromid/kobalt(II)-bromid}},}\ }\href {\doibase
  https://doi.org/10.1016/0022-1902(75)80287-9} {\bibfield  {journal} {\bibinfo
   {journal} {J. Inorg. Nucl. Chem.}\ }\textbf {\bibinfo {volume} {37}},\
  \bibinfo {pages} {1625} (\bibinfo {year} {1975})}\BibitemShut {NoStop}%
\bibitem [{\citenamefont {Seifert}(1977)}]{Seifert_ThAct_1977_A2BX4summary}%
  \BibitemOpen
  \bibfield  {author} {\bibinfo {author} {\bibfnamefont {H.-J.}\ \bibnamefont
  {Seifert}},\ }\bibfield  {title} {\enquote {\bibinfo {title} {{Investigation
  of phase diagrams by DTA and X-ray methods: The systems AX/CoX2 (A = Na-Cs,
  TI;X = Cl, Br, I)}},}\ }\href {\doibase
  https://doi.org/10.1016/0040-6031(77)85037-5} {\bibfield  {journal} {\bibinfo
   {journal} {Thermochim. Acta}\ }\textbf {\bibinfo {volume} {20}},\ \bibinfo
  {pages} {31} (\bibinfo {year} {1977})}\BibitemShut {NoStop}%
\bibitem [{\citenamefont {Schrieffer}\ and\ \citenamefont
  {Wolff}(1966)}]{SchriefferWolff_PR_1966_SchriefferWolff}%
  \BibitemOpen
  \bibfield  {author} {\bibinfo {author} {\bibfnamefont {J.~R.}\ \bibnamefont
  {Schrieffer}}\ and\ \bibinfo {author} {\bibfnamefont {P.~A.}\ \bibnamefont
  {Wolff}},\ }\bibfield  {title} {\enquote {\bibinfo {title} {{Relation between
  the Anderson and Kondo Hamiltonians}},}\ }\href {\doibase
  10.1103/PhysRev.149.491} {\bibfield  {journal} {\bibinfo  {journal} {Phys.
  Rev.}\ }\textbf {\bibinfo {volume} {149}},\ \bibinfo {pages} {491} (\bibinfo
  {year} {1966})}\BibitemShut {NoStop}%
\bibitem [{\citenamefont {Bravyi}\ \emph {et~al.}(2011)\citenamefont {Bravyi},
  \citenamefont {DiVincenzo},\ and\ \citenamefont
  {Loss}}]{Bravyi_AnnPhys_2011_SchriefferWolffReview}%
  \BibitemOpen
  \bibfield  {author} {\bibinfo {author} {\bibfnamefont {S.}~\bibnamefont
  {Bravyi}}, \bibinfo {author} {\bibfnamefont {D.~P.}\ \bibnamefont
  {DiVincenzo}}, \ and\ \bibinfo {author} {\bibfnamefont {D.}~\bibnamefont
  {Loss}},\ }\bibfield  {title} {\enquote {\bibinfo {title} {{Schrieffer-Wolff
  transformation for quantum many-body systems}},}\ }\href {\doibase
  https://doi.org/10.1016/j.aop.2011.06.004} {\bibfield  {journal} {\bibinfo
  {journal} {Ann. Phys.}\ }\textbf {\bibinfo {volume} {326}},\ \bibinfo {pages}
  {2793} (\bibinfo {year} {2011})}\BibitemShut {NoStop}%
\bibitem [{\citenamefont {Scheie}(2018)}]{Scheie_JLTP_2018_LongHeatPulse}%
  \BibitemOpen
  \bibfield  {author} {\bibinfo {author} {\bibfnamefont {A.}~\bibnamefont
  {Scheie}},\ }\bibfield  {title} {\enquote {\bibinfo {title} {{LongHCPulse:
  Long-Pulse Heat Capacity on a Quantum Design PPMS}},}\ }\href {\doibase
  10.1007/s10909-018-2042-9} {\bibfield  {journal} {\bibinfo  {journal} {J. Low
  Temp. Phys.}\ }\textbf {\bibinfo {volume} {193}},\ \bibinfo {pages} {60}
  (\bibinfo {year} {2018})}\BibitemShut {NoStop}%
\bibitem [{\citenamefont {Feng}\ \emph {et~al.}(2018)\citenamefont {Feng},
  \citenamefont {Povarov},\ and\ \citenamefont
  {Zheludev}}]{FengPovarov_PRB_2018_LinariteTilted}%
  \BibitemOpen
  \bibfield  {author} {\bibinfo {author} {\bibfnamefont {Y.}~\bibnamefont
  {Feng}}, \bibinfo {author} {\bibfnamefont {{\relax K. Yu.}}~\bibnamefont
  {Povarov}}, \ and\ \bibinfo {author} {\bibfnamefont {A.}~\bibnamefont
  {Zheludev}},\ }\bibfield  {title} {\enquote {\bibinfo {title} {{Magnetic
  phase diagram of the strongly frustrated quantum spin chain system
  ${\mathrm{PbCuSO}}_{4}{(\mathrm{OH})}_{2}$ in tilted magnetic fields}},}\
  }\href {\doibase 10.1103/PhysRevB.98.054419} {\bibfield  {journal} {\bibinfo
  {journal} {Phys. Rev. B}\ }\textbf {\bibinfo {volume} {98}},\ \bibinfo
  {pages} {054419} (\bibinfo {year} {2018})}\BibitemShut {NoStop}%
\bibitem [{\citenamefont {Blosser}\ \emph {et~al.}(2020)\citenamefont
  {Blosser}, \citenamefont {Facheris},\ and\ \citenamefont
  {Zheludev}}]{Blosser_RevSciInstr_2020_FaradayBalance}%
  \BibitemOpen
  \bibfield  {author} {\bibinfo {author} {\bibfnamefont {D.}~\bibnamefont
  {Blosser}}, \bibinfo {author} {\bibfnamefont {L.}~\bibnamefont {Facheris}}, \
  and\ \bibinfo {author} {\bibfnamefont {A.}~\bibnamefont {Zheludev}},\
  }\bibfield  {title} {\enquote {\bibinfo {title} {{Miniature capacitive
  Faraday force magnetometer for magnetization measurements at low temperatures
  and high magnetic fields}},}\ }\href {\doibase 10.1063/5.0005850} {\bibfield
  {journal} {\bibinfo  {journal} {Rev. Sci. Instr.}\ }\textbf {\bibinfo
  {volume} {91}},\ \bibinfo {pages} {073905} (\bibinfo {year}
  {2020})}\BibitemShut {NoStop}%
\bibitem [{\citenamefont {Henley}(1989)}]{Henley_PRL_1989_OrderByDisorder}%
  \BibitemOpen
  \bibfield  {author} {\bibinfo {author} {\bibfnamefont {C.~L.}\ \bibnamefont
  {Henley}},\ }\bibfield  {title} {\enquote {\bibinfo {title} {Ordering due to
  disorder in a frustrated vector antiferromagnet},}\ }\href {\doibase
  10.1103/PhysRevLett.62.2056} {\bibfield  {journal} {\bibinfo  {journal}
  {Phys. Rev. Lett.}\ }\textbf {\bibinfo {volume} {62}},\ \bibinfo {pages}
  {2056--2059} (\bibinfo {year} {1989})}\BibitemShut {NoStop}%
\bibitem [{\citenamefont {Chubukov}\ and\ \citenamefont
  {Golosov}(1991)}]{ChubukovGolosov_JPCM_1991_QuantumUUD}%
  \BibitemOpen
  \bibfield  {author} {\bibinfo {author} {\bibfnamefont {A.~V.}\ \bibnamefont
  {Chubukov}}\ and\ \bibinfo {author} {\bibfnamefont {D.~I.}\ \bibnamefont
  {Golosov}},\ }\bibfield  {title} {\enquote {\bibinfo {title} {{Quantum theory
  of an antiferromagnet on a triangular lattice in a magnetic field}},}\ }\href
  {\doibase 10.1088/0953-8984/3/1/005} {\bibfield  {journal} {\bibinfo
  {journal} {J. Phys.: Cond. Mat.}\ }\textbf {\bibinfo {volume} {3}},\ \bibinfo
  {pages} {69--82} (\bibinfo {year} {1991})}\BibitemShut {NoStop}%
\bibitem [{\citenamefont {Ranjith}\ \emph
  {et~al.}(2019{\natexlab{a}})\citenamefont {Ranjith}, \citenamefont
  {Dmytriieva}, \citenamefont {Khim}, \citenamefont {Sichelschmidt},
  \citenamefont {Luther}, \citenamefont {Ehlers}, \citenamefont {Yasuoka},
  \citenamefont {Wosnitza}, \citenamefont {Tsirlin}, \citenamefont {K\"uhne},\
  and\ \citenamefont {Baenitz}}]{RanjithDmytriieva_PRB_2019_NaYbO2first}%
  \BibitemOpen
  \bibfield  {author} {\bibinfo {author} {\bibfnamefont {K.~M.}\ \bibnamefont
  {Ranjith}}, \bibinfo {author} {\bibfnamefont {D.}~\bibnamefont {Dmytriieva}},
  \bibinfo {author} {\bibfnamefont {S.}~\bibnamefont {Khim}}, \bibinfo {author}
  {\bibfnamefont {J.}~\bibnamefont {Sichelschmidt}}, \bibinfo {author}
  {\bibfnamefont {S.}~\bibnamefont {Luther}}, \bibinfo {author} {\bibfnamefont
  {D.}~\bibnamefont {Ehlers}}, \bibinfo {author} {\bibfnamefont
  {H.}~\bibnamefont {Yasuoka}}, \bibinfo {author} {\bibfnamefont
  {J.}~\bibnamefont {Wosnitza}}, \bibinfo {author} {\bibfnamefont {A.~A.}\
  \bibnamefont {Tsirlin}}, \bibinfo {author} {\bibfnamefont {H.}~\bibnamefont
  {K\"uhne}}, \ and\ \bibinfo {author} {\bibfnamefont {M.}~\bibnamefont
  {Baenitz}},\ }\bibfield  {title} {\enquote {\bibinfo {title} {{Field-induced
  instability of the quantum spin liquid ground state in the
  ${J}_{\mathrm{eff}}=\frac{1}{2}$ triangular-lattice compound
  ${\mathrm{NaYbO}}_{2}$}},}\ }\href {\doibase 10.1103/PhysRevB.99.180401}
  {\bibfield  {journal} {\bibinfo  {journal} {Phys. Rev. B}\ }\textbf {\bibinfo
  {volume} {99}},\ \bibinfo {pages} {180401} (\bibinfo {year}
  {2019}{\natexlab{a}})}\BibitemShut {NoStop}%
\bibitem [{\citenamefont {Ding}\ \emph {et~al.}(2019)\citenamefont {Ding},
  \citenamefont {Manuel}, \citenamefont {Bachus}, \citenamefont {Gru\ss{}ler},
  \citenamefont {Gegenwart}, \citenamefont {Singleton}, \citenamefont
  {Johnson}, \citenamefont {Walker}, \citenamefont {Adroja}, \citenamefont
  {Hillier},\ and\ \citenamefont
  {Tsirlin}}]{DingManuel_PRB_2019_NaYbO2gaplessproposal}%
  \BibitemOpen
  \bibfield  {author} {\bibinfo {author} {\bibfnamefont {L.}~\bibnamefont
  {Ding}}, \bibinfo {author} {\bibfnamefont {P.}~\bibnamefont {Manuel}},
  \bibinfo {author} {\bibfnamefont {S.}~\bibnamefont {Bachus}}, \bibinfo
  {author} {\bibfnamefont {F.}~\bibnamefont {Gru\ss{}ler}}, \bibinfo {author}
  {\bibfnamefont {P.}~\bibnamefont {Gegenwart}}, \bibinfo {author}
  {\bibfnamefont {J.}~\bibnamefont {Singleton}}, \bibinfo {author}
  {\bibfnamefont {R.~D.}\ \bibnamefont {Johnson}}, \bibinfo {author}
  {\bibfnamefont {H.~C.}\ \bibnamefont {Walker}}, \bibinfo {author}
  {\bibfnamefont {D.~T.}\ \bibnamefont {Adroja}}, \bibinfo {author}
  {\bibfnamefont {A.~D.}\ \bibnamefont {Hillier}}, \ and\ \bibinfo {author}
  {\bibfnamefont {A.~A.}\ \bibnamefont {Tsirlin}},\ }\bibfield  {title}
  {\enquote {\bibinfo {title} {{Gapless spin-liquid state in the structurally
  disorder-free triangular antiferromagnet ${\mathrm{NaYbO}}_{2}$}},}\ }\href
  {\doibase 10.1103/PhysRevB.100.144432} {\bibfield  {journal} {\bibinfo
  {journal} {Phys. Rev. B}\ }\textbf {\bibinfo {volume} {100}},\ \bibinfo
  {pages} {144432} (\bibinfo {year} {2019})}\BibitemShut {NoStop}%
\bibitem [{\citenamefont {Bordelon}\ \emph {et~al.}(2019)\citenamefont
  {Bordelon}, \citenamefont {Kenney}, \citenamefont {Liu}, \citenamefont
  {Hogan}, \citenamefont {Posthuma}, \citenamefont {Kavand}, \citenamefont
  {Lyu}, \citenamefont {Sherwin}, \citenamefont {Butch}, \citenamefont {Brown},
  \citenamefont {Graf}, \citenamefont {Balents},\ and\ \citenamefont
  {Wilson}}]{Bordelon_NatPhys_2019_NaYbO2uud}%
  \BibitemOpen
  \bibfield  {author} {\bibinfo {author} {\bibfnamefont {M.~M.}\ \bibnamefont
  {Bordelon}}, \bibinfo {author} {\bibfnamefont {E.}~\bibnamefont {Kenney}},
  \bibinfo {author} {\bibfnamefont {C.}~\bibnamefont {Liu}}, \bibinfo {author}
  {\bibfnamefont {T.}~\bibnamefont {Hogan}}, \bibinfo {author} {\bibfnamefont
  {L.}~\bibnamefont {Posthuma}}, \bibinfo {author} {\bibfnamefont
  {M.}~\bibnamefont {Kavand}}, \bibinfo {author} {\bibfnamefont
  {Y.}~\bibnamefont {Lyu}}, \bibinfo {author} {\bibfnamefont {M.}~\bibnamefont
  {Sherwin}}, \bibinfo {author} {\bibfnamefont {N.~P.}\ \bibnamefont {Butch}},
  \bibinfo {author} {\bibfnamefont {C.}~\bibnamefont {Brown}}, \bibinfo
  {author} {\bibfnamefont {M.~J.}\ \bibnamefont {Graf}}, \bibinfo {author}
  {\bibfnamefont {L.}~\bibnamefont {Balents}}, \ and\ \bibinfo {author}
  {\bibfnamefont {S.~D.}\ \bibnamefont {Wilson}},\ }\bibfield  {title}
  {\enquote {\bibinfo {title} {{Field-tunable quantum disordered ground state
  in the triangular-lattice antiferromagnet NaYbO$_2$}},}\ }\href {\doibase
  10.1038/s41567-019-0594-5} {\bibfield  {journal} {\bibinfo  {journal} {Nat.
  Phys.}\ }\textbf {\bibinfo {volume} {15}},\ \bibinfo {pages} {1058} (\bibinfo
  {year} {2019})}\BibitemShut {NoStop}%
\bibitem [{\citenamefont {Ranjith}\ \emph
  {et~al.}(2019{\natexlab{b}})\citenamefont {Ranjith}, \citenamefont {Luther},
  \citenamefont {Reimann}, \citenamefont {Schmidt}, \citenamefont {Schlender},
  \citenamefont {Sichelschmidt}, \citenamefont {Yasuoka}, \citenamefont
  {Strydom}, \citenamefont {Skourski}, \citenamefont {Wosnitza}, \citenamefont
  {K\"uhne}, \citenamefont {Doert},\ and\ \citenamefont
  {Baenitz}}]{RanjithLuther_PRB_2019_NaYbSe2uud}%
  \BibitemOpen
  \bibfield  {author} {\bibinfo {author} {\bibfnamefont {K.~M.}\ \bibnamefont
  {Ranjith}}, \bibinfo {author} {\bibfnamefont {S.}~\bibnamefont {Luther}},
  \bibinfo {author} {\bibfnamefont {T.}~\bibnamefont {Reimann}}, \bibinfo
  {author} {\bibfnamefont {B.}~\bibnamefont {Schmidt}}, \bibinfo {author}
  {\bibfnamefont {Ph.}\ \bibnamefont {Schlender}}, \bibinfo {author}
  {\bibfnamefont {J.}~\bibnamefont {Sichelschmidt}}, \bibinfo {author}
  {\bibfnamefont {H.}~\bibnamefont {Yasuoka}}, \bibinfo {author} {\bibfnamefont
  {A.~M.}\ \bibnamefont {Strydom}}, \bibinfo {author} {\bibfnamefont
  {Y.}~\bibnamefont {Skourski}}, \bibinfo {author} {\bibfnamefont
  {J.}~\bibnamefont {Wosnitza}}, \bibinfo {author} {\bibfnamefont
  {H.}~\bibnamefont {K\"uhne}}, \bibinfo {author} {\bibfnamefont {Th.}\
  \bibnamefont {Doert}}, \ and\ \bibinfo {author} {\bibfnamefont
  {M.}~\bibnamefont {Baenitz}},\ }\bibfield  {title} {\enquote {\bibinfo
  {title} {{Anisotropic field-induced ordering in the triangular-lattice
  quantum spin liquid ${\mathrm{NaYbSe}}_{2}$}},}\ }\href {\doibase
  10.1103/PhysRevB.100.224417} {\bibfield  {journal} {\bibinfo  {journal}
  {Phys. Rev. B}\ }\textbf {\bibinfo {volume} {100}},\ \bibinfo {pages}
  {224417} (\bibinfo {year} {2019}{\natexlab{b}})}\BibitemShut {NoStop}%
\bibitem [{\citenamefont {Baenitz}\ \emph {et~al.}(2018)\citenamefont
  {Baenitz}, \citenamefont {Schlender}, \citenamefont {Sichelschmidt},
  \citenamefont {Onykiienko}, \citenamefont {Zangeneh}, \citenamefont
  {Ranjith}, \citenamefont {Sarkar}, \citenamefont {Hozoi}, \citenamefont
  {Walker}, \citenamefont {Orain}, \citenamefont {Yasuoka}, \citenamefont
  {van~den Brink}, \citenamefont {Klauss}, \citenamefont {Inosov},\ and\
  \citenamefont {Doert}}]{BaenitzSchlender_PRB_2018_NaYbS2first}%
  \BibitemOpen
  \bibfield  {author} {\bibinfo {author} {\bibfnamefont {M.}~\bibnamefont
  {Baenitz}}, \bibinfo {author} {\bibfnamefont {Ph.}\ \bibnamefont
  {Schlender}}, \bibinfo {author} {\bibfnamefont {J.}~\bibnamefont
  {Sichelschmidt}}, \bibinfo {author} {\bibfnamefont {Y.~A.}\ \bibnamefont
  {Onykiienko}}, \bibinfo {author} {\bibfnamefont {Z.}~\bibnamefont
  {Zangeneh}}, \bibinfo {author} {\bibfnamefont {K.~M.}\ \bibnamefont
  {Ranjith}}, \bibinfo {author} {\bibfnamefont {R.}~\bibnamefont {Sarkar}},
  \bibinfo {author} {\bibfnamefont {L.}~\bibnamefont {Hozoi}}, \bibinfo
  {author} {\bibfnamefont {H.~C.}\ \bibnamefont {Walker}}, \bibinfo {author}
  {\bibfnamefont {J.-C.}\ \bibnamefont {Orain}}, \bibinfo {author}
  {\bibfnamefont {H.}~\bibnamefont {Yasuoka}}, \bibinfo {author} {\bibfnamefont
  {J.}~\bibnamefont {van~den Brink}}, \bibinfo {author} {\bibfnamefont {H.~H.}\
  \bibnamefont {Klauss}}, \bibinfo {author} {\bibfnamefont {D.~S.}\
  \bibnamefont {Inosov}}, \ and\ \bibinfo {author} {\bibfnamefont {Th.}\
  \bibnamefont {Doert}},\ }\bibfield  {title} {\enquote {\bibinfo {title}
  {{${\mathrm{NaYbS}}_{2}$: A planar spin-$\frac{1}{2}$ triangular-lattice
  magnet and putative spin liquid}},}\ }\href {\doibase
  10.1103/PhysRevB.98.220409} {\bibfield  {journal} {\bibinfo  {journal} {Phys.
  Rev. B}\ }\textbf {\bibinfo {volume} {98}},\ \bibinfo {pages} {220409}
  (\bibinfo {year} {2018})}\BibitemShut {NoStop}%
\bibitem [{\citenamefont {Ma}\ \emph {et~al.}(2020)\citenamefont {Ma},
  \citenamefont {Li}, \citenamefont {Gao}, \citenamefont {Liu}, \citenamefont
  {Ren}, \citenamefont {Zhang}, \citenamefont {Wang}, \citenamefont {Chen},
  \citenamefont {Embs}, \citenamefont {Feng}, \citenamefont {Zhu},
  \citenamefont {Q.}, \citenamefont {Xiang}, \citenamefont {Chen},
  \citenamefont {Choi}, \citenamefont {Qu}, \citenamefont {Li}, \citenamefont
  {Wang}, \citenamefont {Zhou}, \citenamefont {Su}, \citenamefont {Wang},
  \citenamefont {Zhang},\ and\ \citenamefont
  {Chen}}]{MaLiGao_arXiv_2020_NaYbS2uud}%
  \BibitemOpen
  \bibfield  {author} {\bibinfo {author} {\bibfnamefont {J.}~\bibnamefont
  {Ma}}, \bibinfo {author} {\bibfnamefont {J.}~\bibnamefont {Li}}, \bibinfo
  {author} {\bibfnamefont {Y.~H.}\ \bibnamefont {Gao}}, \bibinfo {author}
  {\bibfnamefont {C.}~\bibnamefont {Liu}}, \bibinfo {author} {\bibfnamefont
  {Q.}~\bibnamefont {Ren}}, \bibinfo {author} {\bibfnamefont {Z.}~\bibnamefont
  {Zhang}}, \bibinfo {author} {\bibfnamefont {Z.}~\bibnamefont {Wang}},
  \bibinfo {author} {\bibfnamefont {R.}~\bibnamefont {Chen}}, \bibinfo {author}
  {\bibfnamefont {J.}~\bibnamefont {Embs}}, \bibinfo {author} {\bibfnamefont
  {E.}~\bibnamefont {Feng}}, \bibinfo {author} {\bibfnamefont {F.}~\bibnamefont
  {Zhu}}, \bibinfo {author} {\bibfnamefont {Huang}\ \bibnamefont {Q.}},
  \bibinfo {author} {\bibfnamefont {Z.}~\bibnamefont {Xiang}}, \bibinfo
  {author} {\bibfnamefont {L.}~\bibnamefont {Chen}}, \bibinfo {author}
  {\bibfnamefont {E.~S.}\ \bibnamefont {Choi}}, \bibinfo {author}
  {\bibfnamefont {Z.}~\bibnamefont {Qu}}, \bibinfo {author} {\bibfnamefont
  {L.}~\bibnamefont {Li}}, \bibinfo {author} {\bibfnamefont {J.}~\bibnamefont
  {Wang}}, \bibinfo {author} {\bibfnamefont {H.}~\bibnamefont {Zhou}}, \bibinfo
  {author} {\bibfnamefont {Y.}~\bibnamefont {Su}}, \bibinfo {author}
  {\bibfnamefont {X.}~\bibnamefont {Wang}}, \bibinfo {author} {\bibfnamefont
  {Q.}~\bibnamefont {Zhang}}, \ and\ \bibinfo {author} {\bibfnamefont
  {G.}~\bibnamefont {Chen}},\ }\bibfield  {title} {\enquote {\bibinfo {title}
  {{Spin-orbit-coupled triangular-lattice spin liquid in rare-earth
  chalcogenides}},}\ }\href {https://arxiv.org/abs/2002.09224} {\bibfield
  {journal} {\bibinfo  {journal} {arXiv}\ ,\ \bibinfo {pages} {2002.09224}}
  (\bibinfo {year} {2020})}\BibitemShut {NoStop}%
\bibitem [{\citenamefont {Chen}\ \emph {et~al.}(2013)\citenamefont {Chen},
  \citenamefont {Ju}, \citenamefont {Jiang}, \citenamefont {Starykh},\ and\
  \citenamefont {Balents}}]{ChenJuJiang_PRB_2013_DistortTriangquasi1D}%
  \BibitemOpen
  \bibfield  {author} {\bibinfo {author} {\bibfnamefont {R.}~\bibnamefont
  {Chen}}, \bibinfo {author} {\bibfnamefont {H.}~\bibnamefont {Ju}}, \bibinfo
  {author} {\bibfnamefont {H.-C.}\ \bibnamefont {Jiang}}, \bibinfo {author}
  {\bibfnamefont {O.~A.}\ \bibnamefont {Starykh}}, \ and\ \bibinfo {author}
  {\bibfnamefont {L.}~\bibnamefont {Balents}},\ }\bibfield  {title} {\enquote
  {\bibinfo {title} {{Ground states of spin-$\frac{1}{2}$ triangular
  antiferromagnets in a magnetic field}},}\ }\href {\doibase
  10.1103/PhysRevB.87.165123} {\bibfield  {journal} {\bibinfo  {journal} {Phys.
  Rev. B}\ }\textbf {\bibinfo {volume} {87}},\ \bibinfo {pages} {165123}
  (\bibinfo {year} {2013})}\BibitemShut {NoStop}%
\bibitem [{\citenamefont {Griset}\ \emph {et~al.}(2011)\citenamefont {Griset},
  \citenamefont {Head}, \citenamefont {Alicea},\ and\ \citenamefont
  {Starykh}}]{GrisetHead_PRB_2011_DefTriangwithDM}%
  \BibitemOpen
  \bibfield  {author} {\bibinfo {author} {\bibfnamefont {C.}~\bibnamefont
  {Griset}}, \bibinfo {author} {\bibfnamefont {S.}~\bibnamefont {Head}},
  \bibinfo {author} {\bibfnamefont {J.}~\bibnamefont {Alicea}}, \ and\ \bibinfo
  {author} {\bibfnamefont {O.~A.}\ \bibnamefont {Starykh}},\ }\bibfield
  {title} {\enquote {\bibinfo {title} {{Deformed triangular lattice
  antiferromagnets in a magnetic field: Role of spatial anisotropy and
  Dzyaloshinskii-Moriya interactions}},}\ }\href {\doibase
  10.1103/PhysRevB.84.245108} {\bibfield  {journal} {\bibinfo  {journal} {Phys.
  Rev. B}\ }\textbf {\bibinfo {volume} {84}},\ \bibinfo {pages} {245108}
  (\bibinfo {year} {2011})}\BibitemShut {NoStop}%
\bibitem [{\citenamefont {Quirion}\ \emph {et~al.}(2015)\citenamefont
  {Quirion}, \citenamefont {Lapointe-Major}, \citenamefont {Poirier},
  \citenamefont {Quilliam}, \citenamefont {Dun},\ and\ \citenamefont
  {Zhou}}]{Quirion_PRB_2015_Ba3CoSb2O9ultrasonicUUD}%
  \BibitemOpen
  \bibfield  {author} {\bibinfo {author} {\bibfnamefont {G.}~\bibnamefont
  {Quirion}}, \bibinfo {author} {\bibfnamefont {M.}~\bibnamefont
  {Lapointe-Major}}, \bibinfo {author} {\bibfnamefont {M.}~\bibnamefont
  {Poirier}}, \bibinfo {author} {\bibfnamefont {J.~A.}\ \bibnamefont
  {Quilliam}}, \bibinfo {author} {\bibfnamefont {Z.~L.}\ \bibnamefont {Dun}}, \
  and\ \bibinfo {author} {\bibfnamefont {H.~D.}\ \bibnamefont {Zhou}},\
  }\bibfield  {title} {\enquote {\bibinfo {title} {{Magnetic phase diagram of
  ${\mathrm{Ba}}_{3}{\mathrm{CoSb}}_{2}{\mathrm{O}}_{9}$ as determined by
  ultrasound velocity measurements}},}\ }\href {\doibase
  10.1103/PhysRevB.92.014414} {\bibfield  {journal} {\bibinfo  {journal} {Phys.
  Rev. B}\ }\textbf {\bibinfo {volume} {92}},\ \bibinfo {pages} {014414}
  (\bibinfo {year} {2015})}\BibitemShut {NoStop}%
\bibitem [{\citenamefont {Koutroulakis}\ \emph {et~al.}(2015)\citenamefont
  {Koutroulakis}, \citenamefont {Zhou}, \citenamefont {Kamiya}, \citenamefont
  {Thompson}, \citenamefont {Zhou}, \citenamefont {Batista},\ and\
  \citenamefont {Brown}}]{KoutroulakisZhou_PRB_2015_Ba3CoSb2O9NMRUUD}%
  \BibitemOpen
  \bibfield  {author} {\bibinfo {author} {\bibfnamefont {G.}~\bibnamefont
  {Koutroulakis}}, \bibinfo {author} {\bibfnamefont {T.}~\bibnamefont {Zhou}},
  \bibinfo {author} {\bibfnamefont {Y.}~\bibnamefont {Kamiya}}, \bibinfo
  {author} {\bibfnamefont {J.~D.}\ \bibnamefont {Thompson}}, \bibinfo {author}
  {\bibfnamefont {H.~D.}\ \bibnamefont {Zhou}}, \bibinfo {author}
  {\bibfnamefont {C.~D.}\ \bibnamefont {Batista}}, \ and\ \bibinfo {author}
  {\bibfnamefont {S.~E.}\ \bibnamefont {Brown}},\ }\bibfield  {title} {\enquote
  {\bibinfo {title} {{Quantum phase diagram of the $S=\frac{1}{2}$
  triangular-lattice antiferromagnet
  ${\mathrm{Ba}}_{3}{\mathrm{CoSb}}_{2}{\mathrm{O}}_{9}$}},}\ }\href {\doibase
  10.1103/PhysRevB.91.024410} {\bibfield  {journal} {\bibinfo  {journal} {Phys.
  Rev. B}\ }\textbf {\bibinfo {volume} {91}},\ \bibinfo {pages} {024410}
  (\bibinfo {year} {2015})}\BibitemShut {NoStop}%
\bibitem [{\citenamefont {Kamiya}\ \emph {et~al.}(2018)\citenamefont {Kamiya},
  \citenamefont {Ge}, \citenamefont {Hong}, \citenamefont {Qiu}, \citenamefont
  {Quintero-Castro}, \citenamefont {Lu}, \citenamefont {Cao}, \citenamefont
  {Matsuda}, \citenamefont {Choi}, \citenamefont {Batista}, \citenamefont
  {Mourigal}, \citenamefont {D.},\ and\ \citenamefont
  {Ma}}]{KamiyaGe_NatComm_2018_Ba3CoSb2O9spectrumUUD}%
  \BibitemOpen
  \bibfield  {author} {\bibinfo {author} {\bibfnamefont {Y.}~\bibnamefont
  {Kamiya}}, \bibinfo {author} {\bibfnamefont {L.}~\bibnamefont {Ge}}, \bibinfo
  {author} {\bibfnamefont {T.}~\bibnamefont {Hong}}, \bibinfo {author}
  {\bibfnamefont {Y.}~\bibnamefont {Qiu}}, \bibinfo {author} {\bibfnamefont
  {D.~L.}\ \bibnamefont {Quintero-Castro}}, \bibinfo {author} {\bibfnamefont
  {Z.}~\bibnamefont {Lu}}, \bibinfo {author} {\bibfnamefont {H.~B.}\
  \bibnamefont {Cao}}, \bibinfo {author} {\bibfnamefont {M.}~\bibnamefont
  {Matsuda}}, \bibinfo {author} {\bibfnamefont {E.~S.}\ \bibnamefont {Choi}},
  \bibinfo {author} {\bibfnamefont {C.~D.}\ \bibnamefont {Batista}}, \bibinfo
  {author} {\bibfnamefont {M.}~\bibnamefont {Mourigal}}, \bibinfo {author}
  {\bibfnamefont {Zhou~H.}\ \bibnamefont {D.}}, \ and\ \bibinfo {author}
  {\bibfnamefont {J.}~\bibnamefont {Ma}},\ }\bibfield  {title} {\enquote
  {\bibinfo {title} {{The nature of spin excitations in the one-third
  magnetization plateau phase of Ba3CoSb2O9}},}\ }\href {\doibase
  10.1038/s41467-018-04914-1} {\bibfield  {journal} {\bibinfo  {journal} {Nat.
  Commun.}\ }\textbf {\bibinfo {volume} {9}},\ \bibinfo {pages} {2666}
  (\bibinfo {year} {2018})}\BibitemShut {NoStop}%
\bibitem [{\citenamefont {Gosuly}(2016)}]{Gosuly_2016_PhDthesis}%
  \BibitemOpen
  \bibfield  {author} {\bibinfo {author} {\bibfnamefont {S.~P.}\ \bibnamefont
  {Gosuly}},\ }\href {http://discovery.ucl.ac.uk/id/eprint/1553138} {\emph
  {\bibinfo {title} {{Neutron Scattering Studies of Low-Dimensional Quantum
  Spin Systems}}}}\ (\bibinfo  {publisher} {PhD thesis, University College
  London},\ \bibinfo {year} {2016})\BibitemShut {NoStop}%
\bibitem [{\citenamefont {Becker}\ \emph {et~al.}(2015)\citenamefont {Becker},
  \citenamefont {Hermanns}, \citenamefont {Bauer}, \citenamefont {Garst},\ and\
  \citenamefont {Trebst}}]{BeckerHermanns_PRB_2015_TriangularKitaevGS}%
  \BibitemOpen
  \bibfield  {author} {\bibinfo {author} {\bibfnamefont {M.}~\bibnamefont
  {Becker}}, \bibinfo {author} {\bibfnamefont {M.}~\bibnamefont {Hermanns}},
  \bibinfo {author} {\bibfnamefont {B.}~\bibnamefont {Bauer}}, \bibinfo
  {author} {\bibfnamefont {M.}~\bibnamefont {Garst}}, \ and\ \bibinfo {author}
  {\bibfnamefont {S.}~\bibnamefont {Trebst}},\ }\bibfield  {title} {\enquote
  {\bibinfo {title} {{Spin-orbit physics of $j=\frac{1}{2}$ Mott insulators on
  the triangular lattice}},}\ }\href {\doibase 10.1103/PhysRevB.91.155135}
  {\bibfield  {journal} {\bibinfo  {journal} {Phys. Rev. B}\ }\textbf {\bibinfo
  {volume} {91}},\ \bibinfo {pages} {155135} (\bibinfo {year}
  {2015})}\BibitemShut {NoStop}%
\bibitem [{\citenamefont {Rousochatzakis}\ \emph {et~al.}(2016)\citenamefont
  {Rousochatzakis}, \citenamefont {R\"ossler}, \citenamefont {van~den Brink},\
  and\ \citenamefont
  {Daghofer}}]{RousochatzakisRossler_PRB_2016_TriangularKitaevGS}%
  \BibitemOpen
  \bibfield  {author} {\bibinfo {author} {\bibfnamefont {I.}~\bibnamefont
  {Rousochatzakis}}, \bibinfo {author} {\bibfnamefont {U.~K.}\ \bibnamefont
  {R\"ossler}}, \bibinfo {author} {\bibfnamefont {J.}~\bibnamefont {van~den
  Brink}}, \ and\ \bibinfo {author} {\bibfnamefont {M.}~\bibnamefont
  {Daghofer}},\ }\bibfield  {title} {\enquote {\bibinfo {title} {Kitaev
  anisotropy induces mesoscopic ${{Z}}_{2}$ vortex crystals in frustrated
  hexagonal antiferromagnets},}\ }\href {\doibase 10.1103/PhysRevB.93.104417}
  {\bibfield  {journal} {\bibinfo  {journal} {Phys. Rev. B}\ }\textbf {\bibinfo
  {volume} {93}},\ \bibinfo {pages} {104417} (\bibinfo {year}
  {2016})}\BibitemShut {NoStop}%
\bibitem [{\citenamefont {Kishimoto}\ \emph {et~al.}(2018)\citenamefont
  {Kishimoto}, \citenamefont {Morita}, \citenamefont {Matsubayashi},
  \citenamefont {Sota}, \citenamefont {Yunoki},\ and\ \citenamefont
  {Tohyama}}]{KishimotoMorita_PRB_2018_HoneyTriangularKitaevGS}%
  \BibitemOpen
  \bibfield  {author} {\bibinfo {author} {\bibfnamefont {M.}~\bibnamefont
  {Kishimoto}}, \bibinfo {author} {\bibfnamefont {K.}~\bibnamefont {Morita}},
  \bibinfo {author} {\bibfnamefont {Y.}~\bibnamefont {Matsubayashi}}, \bibinfo
  {author} {\bibfnamefont {S.}~\bibnamefont {Sota}}, \bibinfo {author}
  {\bibfnamefont {S.}~\bibnamefont {Yunoki}}, \ and\ \bibinfo {author}
  {\bibfnamefont {T.}~\bibnamefont {Tohyama}},\ }\bibfield  {title} {\enquote
  {\bibinfo {title} {{Ground state phase diagram of the Kitaev-Heisenberg model
  on a honeycomb-triangular lattice}},}\ }\href {\doibase
  10.1103/PhysRevB.98.054411} {\bibfield  {journal} {\bibinfo  {journal} {Phys.
  Rev. B}\ }\textbf {\bibinfo {volume} {98}},\ \bibinfo {pages} {054411}
  (\bibinfo {year} {2018})}\BibitemShut {NoStop}%
\bibitem [{\citenamefont {Coldea}\ \emph {et~al.}(2002)\citenamefont {Coldea},
  \citenamefont {Tennant}, \citenamefont {Habicht}, \citenamefont {Smeibidl},
  \citenamefont {Wolters},\ and\ \citenamefont
  {Tylczynski}}]{Coldea_PRL_2002_Cs2CuCl4highfield}%
  \BibitemOpen
  \bibfield  {author} {\bibinfo {author} {\bibfnamefont {R.}~\bibnamefont
  {Coldea}}, \bibinfo {author} {\bibfnamefont {D.~A.}\ \bibnamefont {Tennant}},
  \bibinfo {author} {\bibfnamefont {K.}~\bibnamefont {Habicht}}, \bibinfo
  {author} {\bibfnamefont {P.}~\bibnamefont {Smeibidl}}, \bibinfo {author}
  {\bibfnamefont {C.}~\bibnamefont {Wolters}}, \ and\ \bibinfo {author}
  {\bibfnamefont {Z.}~\bibnamefont {Tylczynski}},\ }\bibfield  {title}
  {\enquote {\bibinfo {title} {{Direct Measurement of the Spin Hamiltonian and
  Observation of Condensation of Magnons in the 2D Frustrated Quantum Magnet
  ${\mathrm{Cs}}_{2}{\mathrm{CuCl}}_{4}$}},}\ }\href {\doibase
  10.1103/PhysRevLett.88.137203} {\bibfield  {journal} {\bibinfo  {journal}
  {Phys. Rev. Lett.}\ }\textbf {\bibinfo {volume} {88}},\ \bibinfo {pages}
  {137203} (\bibinfo {year} {2002})}\BibitemShut {NoStop}%
\bibitem [{\citenamefont {Povarov}\ \emph {et~al.}(2011)\citenamefont
  {Povarov}, \citenamefont {Smirnov}, \citenamefont {Starykh}, \citenamefont
  {Petrov},\ and\ \citenamefont
  {Shapiro}}]{PovarovSmirnov_PRL_2011_ESRdoublet}%
  \BibitemOpen
  \bibfield  {author} {\bibinfo {author} {\bibfnamefont {{\relax K.
  Yu.}}~\bibnamefont {Povarov}}, \bibinfo {author} {\bibfnamefont {A.~I.}\
  \bibnamefont {Smirnov}}, \bibinfo {author} {\bibfnamefont {O.~A.}\
  \bibnamefont {Starykh}}, \bibinfo {author} {\bibfnamefont {S.~V.}\
  \bibnamefont {Petrov}}, \ and\ \bibinfo {author} {\bibfnamefont {{\relax A.
  Ya.}}~\bibnamefont {Shapiro}},\ }\bibfield  {title} {\enquote {\bibinfo
  {title} {{Modes of Magnetic Resonance in the Spin-Liquid Phase of
  ${\mathrm{Cs}}_{2}{\mathrm{CuCl}}_{4}$}},}\ }\href {\doibase
  10.1103/PhysRevLett.107.037204} {\bibfield  {journal} {\bibinfo  {journal}
  {Phys. Rev. Lett.}\ }\textbf {\bibinfo {volume} {107}},\ \bibinfo {pages}
  {037204} (\bibinfo {year} {2011})}\BibitemShut {NoStop}%
\bibitem [{\citenamefont {Sears}\ \emph {et~al.}(2015)\citenamefont {Sears},
  \citenamefont {Songvilay}, \citenamefont {Plumb}, \citenamefont {Clancy},
  \citenamefont {Qiu}, \citenamefont {Zhao}, \citenamefont {Parshall},\ and\
  \citenamefont {Kim}}]{SearsSongvilay_PRB_2015_alphaRuClsusceptible}%
  \BibitemOpen
  \bibfield  {author} {\bibinfo {author} {\bibfnamefont {J.~A.}\ \bibnamefont
  {Sears}}, \bibinfo {author} {\bibfnamefont {M.}~\bibnamefont {Songvilay}},
  \bibinfo {author} {\bibfnamefont {K.~W.}\ \bibnamefont {Plumb}}, \bibinfo
  {author} {\bibfnamefont {J.~P.}\ \bibnamefont {Clancy}}, \bibinfo {author}
  {\bibfnamefont {Y.}~\bibnamefont {Qiu}}, \bibinfo {author} {\bibfnamefont
  {Y.}~\bibnamefont {Zhao}}, \bibinfo {author} {\bibfnamefont {D.}~\bibnamefont
  {Parshall}}, \ and\ \bibinfo {author} {\bibfnamefont {Y.-J.}\ \bibnamefont
  {Kim}},\ }\bibfield  {title} {\enquote {\bibinfo {title} {{Magnetic order in
  $\ensuremath{\alpha}\ensuremath{-}{\text{RuCl}}_{3}$: A honeycomb-lattice
  quantum magnet with strong spin-orbit coupling}},}\ }\href {\doibase
  10.1103/PhysRevB.91.144420} {\bibfield  {journal} {\bibinfo  {journal} {Phys.
  Rev. B}\ }\textbf {\bibinfo {volume} {91}},\ \bibinfo {pages} {144420}
  (\bibinfo {year} {2015})}\BibitemShut {NoStop}%
\bibitem [{\citenamefont {He}\ \emph {et~al.}(2009)\citenamefont {He},
  \citenamefont {Yamaura},\ and\ \citenamefont
  {Cheng}}]{HeYamaura_JACS_2009_CoV2O6plateus}%
  \BibitemOpen
  \bibfield  {author} {\bibinfo {author} {\bibfnamefont {Z.}~\bibnamefont
  {He}}, \bibinfo {author} {\bibfnamefont {Y.}~\bibnamefont {Yamaura},
  \bibfnamefont {J.-I.~Ueda}}, \ and\ \bibinfo {author} {\bibfnamefont
  {W.}~\bibnamefont {Cheng}},\ }\bibfield  {title} {\enquote {\bibinfo {title}
  {{CoV$_2$O$_6$ Single Crystals Grown in a Closed Crucible: Unusual Magnetic
  Behaviors with Large Anisotropy and $1/3$ Magnetization Plateau}},}\ }\href
  {\doibase 10.1021/ja902623b} {\bibfield  {journal} {\bibinfo  {journal} {J.
  Am. Chem. Soc.}\ }\textbf {\bibinfo {volume} {131}},\ \bibinfo {pages} {7554}
  (\bibinfo {year} {2009})}\BibitemShut {NoStop}%
\bibitem [{\citenamefont {Lenertz}\ \emph {et~al.}(2012)\citenamefont
  {Lenertz}, \citenamefont {Alaria}, \citenamefont {Stoeffler}, \citenamefont
  {Colis}, \citenamefont {Dinia}, \citenamefont {Mentr\'e}, \citenamefont
  {Andr\'e}, \citenamefont {Porcher},\ and\ \citenamefont
  {Suard}}]{LenertzAlaria_PRB_2012_CoV2O6groundstates}%
  \BibitemOpen
  \bibfield  {author} {\bibinfo {author} {\bibfnamefont {M.}~\bibnamefont
  {Lenertz}}, \bibinfo {author} {\bibfnamefont {J.}~\bibnamefont {Alaria}},
  \bibinfo {author} {\bibfnamefont {D.}~\bibnamefont {Stoeffler}}, \bibinfo
  {author} {\bibfnamefont {S.}~\bibnamefont {Colis}}, \bibinfo {author}
  {\bibfnamefont {A.}~\bibnamefont {Dinia}}, \bibinfo {author} {\bibfnamefont
  {O.}~\bibnamefont {Mentr\'e}}, \bibinfo {author} {\bibfnamefont
  {G.}~\bibnamefont {Andr\'e}}, \bibinfo {author} {\bibfnamefont
  {F.}~\bibnamefont {Porcher}}, \ and\ \bibinfo {author} {\bibfnamefont
  {E.}~\bibnamefont {Suard}},\ }\bibfield  {title} {\enquote {\bibinfo {title}
  {{Magnetic structure of ground and field-induced ordered states of
  low-dimensional $\ensuremath{\alpha}$-CoV${}_{2}$O${}_{6}$: Experiment and
  theory}},}\ }\href {\doibase 10.1103/PhysRevB.86.214428} {\bibfield
  {journal} {\bibinfo  {journal} {Phys. Rev. B}\ }\textbf {\bibinfo {volume}
  {86}},\ \bibinfo {pages} {214428} (\bibinfo {year} {2012})}\BibitemShut
  {NoStop}%
\bibitem [{\citenamefont {Markkula}\ \emph {et~al.}(2012)\citenamefont
  {Markkula}, \citenamefont {Ar\'{e}valo-L\'{o}pez},\ and\ \citenamefont
  {Attfield}}]{Markkula_PRB_2012_CoV2O6groundstates2}%
  \BibitemOpen
  \bibfield  {author} {\bibinfo {author} {\bibfnamefont {M.}~\bibnamefont
  {Markkula}}, \bibinfo {author} {\bibfnamefont {A.~M.}\ \bibnamefont
  {Ar\'{e}valo-L\'{o}pez}}, \ and\ \bibinfo {author} {\bibfnamefont {J.~P.}\
  \bibnamefont {Attfield}},\ }\bibfield  {title} {\enquote {\bibinfo {title}
  {{Field-induced spin orders in monoclinic CoV${}_{2}$O${}_{6}$}},}\ }\href
  {\doibase 10.1103/PhysRevB.86.134401} {\bibfield  {journal} {\bibinfo
  {journal} {Phys. Rev. B}\ }\textbf {\bibinfo {volume} {86}},\ \bibinfo
  {pages} {134401} (\bibinfo {year} {2012})}\BibitemShut {NoStop}%
\bibitem [{\citenamefont {Kim}\ \emph {et~al.}(2012)\citenamefont {Kim},
  \citenamefont {Kim}, \citenamefont {Kim}, \citenamefont {Choi}, \citenamefont
  {Park}, \citenamefont {Jeong},\ and\ \citenamefont
  {Min}}]{KimKimKim_PRB_2012_CoV2O6groundstates3}%
  \BibitemOpen
  \bibfield  {author} {\bibinfo {author} {\bibfnamefont {B.}~\bibnamefont
  {Kim}}, \bibinfo {author} {\bibfnamefont {B.~H.}\ \bibnamefont {Kim}},
  \bibinfo {author} {\bibfnamefont {K.}~\bibnamefont {Kim}}, \bibinfo {author}
  {\bibfnamefont {H.~C.}\ \bibnamefont {Choi}}, \bibinfo {author}
  {\bibfnamefont {S.-Y.}\ \bibnamefont {Park}}, \bibinfo {author}
  {\bibfnamefont {Y.~H.}\ \bibnamefont {Jeong}}, \ and\ \bibinfo {author}
  {\bibfnamefont {B.~I.}\ \bibnamefont {Min}},\ }\bibfield  {title} {\enquote
  {\bibinfo {title} {{Unusual magnetic properties induced by local structure in
  a quasi-one-dimensional Ising chain system:
  $\ensuremath{\alpha}$-CoV${}_{2}$O${}_{6}$}},}\ }\href {\doibase
  10.1103/PhysRevB.85.220407} {\bibfield  {journal} {\bibinfo  {journal} {Phys.
  Rev. B}\ }\textbf {\bibinfo {volume} {85}},\ \bibinfo {pages} {220407}
  (\bibinfo {year} {2012})}\BibitemShut {NoStop}%
\bibitem [{\citenamefont {Sa\'{u}l}\ \emph {et~al.}(2013)\citenamefont
  {Sa\'{u}l}, \citenamefont {Vodenicarevic},\ and\ \citenamefont
  {Radtke}}]{SaulVodenicarevic_PRB_2013_CoV2O6theory}%
  \BibitemOpen
  \bibfield  {author} {\bibinfo {author} {\bibfnamefont {A.}~\bibnamefont
  {Sa\'{u}l}}, \bibinfo {author} {\bibfnamefont {D.}~\bibnamefont
  {Vodenicarevic}}, \ and\ \bibinfo {author} {\bibfnamefont {G.}~\bibnamefont
  {Radtke}},\ }\bibfield  {title} {\enquote {\bibinfo {title} {{Theoretical
  study of the magnetic order in
  $\ensuremath{\alpha}$-CoV${}_{2}$O${}_{6}$}},}\ }\href {\doibase
  10.1103/PhysRevB.87.024403} {\bibfield  {journal} {\bibinfo  {journal} {Phys.
  Rev. B}\ }\textbf {\bibinfo {volume} {87}},\ \bibinfo {pages} {024403}
  (\bibinfo {year} {2013})}\BibitemShut {NoStop}%
\bibitem [{\citenamefont {Edwards}\ \emph {et~al.}(2020)\citenamefont
  {Edwards}, \citenamefont {Lane}, \citenamefont {Wallington}, \citenamefont
  {Arevalo-Lopez}, \citenamefont {Songvilay}, \citenamefont {Pachoud},
  \citenamefont {Niedermayer}, \citenamefont {Tucker}, \citenamefont {Manuel},
  \citenamefont {Paulsen}, \citenamefont {Lhotel}, \citenamefont {Attfield},
  \citenamefont {Giblin},\ and\ \citenamefont
  {Stock}}]{EdwardsLane_PRB_2020_CoV2O6metastable}%
  \BibitemOpen
  \bibfield  {author} {\bibinfo {author} {\bibfnamefont {L.}~\bibnamefont
  {Edwards}}, \bibinfo {author} {\bibfnamefont {H.}~\bibnamefont {Lane}},
  \bibinfo {author} {\bibfnamefont {F.}~\bibnamefont {Wallington}}, \bibinfo
  {author} {\bibfnamefont {A.~M.}\ \bibnamefont {Arevalo-Lopez}}, \bibinfo
  {author} {\bibfnamefont {M.}~\bibnamefont {Songvilay}}, \bibinfo {author}
  {\bibfnamefont {E.}~\bibnamefont {Pachoud}}, \bibinfo {author} {\bibfnamefont
  {Ch.}\ \bibnamefont {Niedermayer}}, \bibinfo {author} {\bibfnamefont
  {G.}~\bibnamefont {Tucker}}, \bibinfo {author} {\bibfnamefont
  {P.}~\bibnamefont {Manuel}}, \bibinfo {author} {\bibfnamefont
  {C.}~\bibnamefont {Paulsen}}, \bibinfo {author} {\bibfnamefont
  {E.}~\bibnamefont {Lhotel}}, \bibinfo {author} {\bibfnamefont {J.~P.}\
  \bibnamefont {Attfield}}, \bibinfo {author} {\bibfnamefont {S.~R.}\
  \bibnamefont {Giblin}}, \ and\ \bibinfo {author} {\bibfnamefont
  {C.}~\bibnamefont {Stock}},\ }\bibfield  {title} {\enquote {\bibinfo {title}
  {{Metastable and localized Ising magnetism in
  $\ensuremath{\alpha}\ensuremath{-}{\mathrm{CoV}}_{2}{\mathrm{O}}_{6}$
  magnetization plateaus}},}\ }\href {\doibase 10.1103/PhysRevB.102.195136}
  {\bibfield  {journal} {\bibinfo  {journal} {Phys. Rev. B}\ }\textbf {\bibinfo
  {volume} {102}},\ \bibinfo {pages} {195136} (\bibinfo {year}
  {2020})}\BibitemShut {NoStop}%
\end{thebibliography}%

\end{document}